\documentclass[logo,address,copyright,10pt]{fdtn}

\usepackage[numbers]{natbib}
\usepackage{booktabs}
\usepackage{multirow}
\tcbuselibrary{most}
\usepackage{listings}
\usepackage{xcolor}
\usepackage{float}
\usepackage{afterpage}
\usepackage{tikz}
\usetikzlibrary{arrows.meta,fit,positioning}

\definecolor{customorange}{HTML}{FF7F0E}
\definecolor{customblue}{HTML}{1F77B4}
\newtcbox{\highlight}[1][customorange]{
    on line,
    arc=0pt,
    colback=#1!35!white,
    colframe=#1!35!white,
    boxsep=0pt,
    left=1pt,
    right=1pt,
    top=2pt,
    bottom=2pt,
    boxrule=0pt,
    nobeforeafter
}

\DeclareFontFamily{OMS}{cmsy}{\skewchar\font48}
\DeclareFontShape{OMS}{cmsy}{m}{n}{%
  <-6>cmsy5%
  <6-7>cmsy6%
  <7-8>cmsy7%
  <8-9>cmsy8%
  <9-10>cmsy9%
  <10->cmsy10%
}{}
\DeclareFontShape{OMS}{cmsy}{b}{n}{%
  <-6>cmbsy5%
  <6-7>cmbsy6%
  <7-8>cmbsy7%
  <8-9>cmbsy8%
  <9-10>cmbsy9%
  <10->cmbsy10%
}{}
\DeclareSymbolFont{cmsymbols}{OMS}{cmsy}{m}{n}
\SetSymbolFont{cmsymbols}{bold}{OMS}{cmsy}{b}{n}
\DeclareMathSymbol{\pm}{\mathbin}{cmsymbols}{"06}

\AtBeginDocument{%
  \count255=\symcmsymbols
  \multiply\count255 by 256
  \advance\count255 by "2000
  \mathcode`\-=\count255
}

\definecolor{codebg}{gray}{0.95}
\newcommand{\code}[1]{\lstinline|#1|}
\newcommand{\FAPO}{FAPO}
\newcommand{\ClaudeCode}{Claude Code}
\newcommand{\GPTFourOneMini}{GPT-4.1-mini}
\newcommand{\GPTFiveFourMini}{GPT-5.4-mini}
\newcommand{\GPTFive}{GPT-5}
\newcommand{\GemmaThreeTwelveB}{Gemma~3-12B}
\newcommand{\ClaudeOpusFourSix}{Claude Opus~4.6}
\newcommand{\FoundationSecEightBInstruct}{Foundation-Sec-8B-Instruct}
\newcommand{\FoundationSecEightBReasoning}{Foundation-Sec-8B-Reasoning}
\newcommand{\FoundationSecEightBInstructShort}{Foundation-Sec-8B-Inst.}
\newcommand{\FoundationSecEightBReasoningShort}{Foundation-Sec-8B-Reas.}
\definecolor{teal}{HTML}{008080}

\title{\FAPO{}: Fully Automated Prompt Optimization\\of Multi-Step LLM Pipelines}

\author{
    Paul~Kassianik\textsuperscript{1,*},
    Baturay~Saglam\textsuperscript{1,2,*,\textdagger},
    Huaibo Zhao\textsuperscript{1},
    Blaine~Nelson\textsuperscript{1},
    Supriti~Vijay\textsuperscript{1},
    Aman~Priyanshu\textsuperscript{1},
    Amin~Karbasi\textsuperscript{1}
    \par\vspace{1.5em}%
    {\Affilfont\setlength{\parskip}{0pt}\setlength{\parindent}{0pt}%
       \textsuperscript{1}Foundation AI--Cisco Systems Inc.\par
       \textsuperscript{2}Yale University\par
    }%
    \par\vspace{1em}%
    {\small\setlength{\parskip}{0pt}\setlength{\parindent}{0pt}%
       \textsuperscript{*}Equal contribution. Corresponding authors: \texttt{\{\href{mailto:basaglam@cisco.com}{basaglam}, \href{mailto:huaibzha@cisco.com}{huaibzha}\}@cisco.com}\par
       \textsuperscript{\textdagger}Work done during an internship at Foundation AI.\par
    }%
}

\date{}

\begin{document}

\begin{abstract}
Multi-step LLM pipelines fail through interactions among retrieval, reasoning, and formatting steps, so prompt-only optimization can miss bottlenecks in the chain.
We present \textit{Fully Automated Prompt Optimization} (\FAPO{}), a framework that lets \ClaudeCode{} optimize an LLM pipeline inside a standardized codebase.
\FAPO{} evaluates a pipeline, inspects intermediate steps, diagnoses failures, proposes scoped changes, and validates variants repeatedly to optimize against a score function.
It first tries prompt edits and, only when prompt optimization appears insufficient, changes chain structure within the permitted scope when attribution identifies a structural bottleneck.
Across six benchmarks and three task models, \FAPO{} beats the baseline GEPA in \textbf{15 of 18} model--benchmark comparisons.
In 11 model--benchmark comparisons, \FAPO{} wins with non-overlapping mean $\pm$ trial-standard-deviation ranges, and the mean \FAPO{}--GEPA gain is \textbf{+14.1\,pp}.
In the six HoVer and IFBench comparisons where prompt-first search escalated to structural changes, \FAPO{} wins all six with a mean gain of \textbf{+33.8\,pp}.
\FAPO{} also improves performance on security tasks: on CTIBench-RCM, a security CVE-to-CWE task, prompt-only \FAPO{} lifts test accuracy by \textbf{+4.0\,pp} on \GPTFive{}, \textbf{+7.1\,pp} on \FoundationSecEightBInstruct{}, and \textbf{+2.0\,pp} on \FoundationSecEightBReasoning{}.
These results position \FAPO{} as a state-of-the-art pipeline optimization technique for both general-purpose and security-focused tasks.
\end{abstract}

\maketitle

\afterpage{%
\begin{figure}[H]
    \centering
    \makebox[\textwidth][c]{%
        \includegraphics[width=1.15\textwidth]{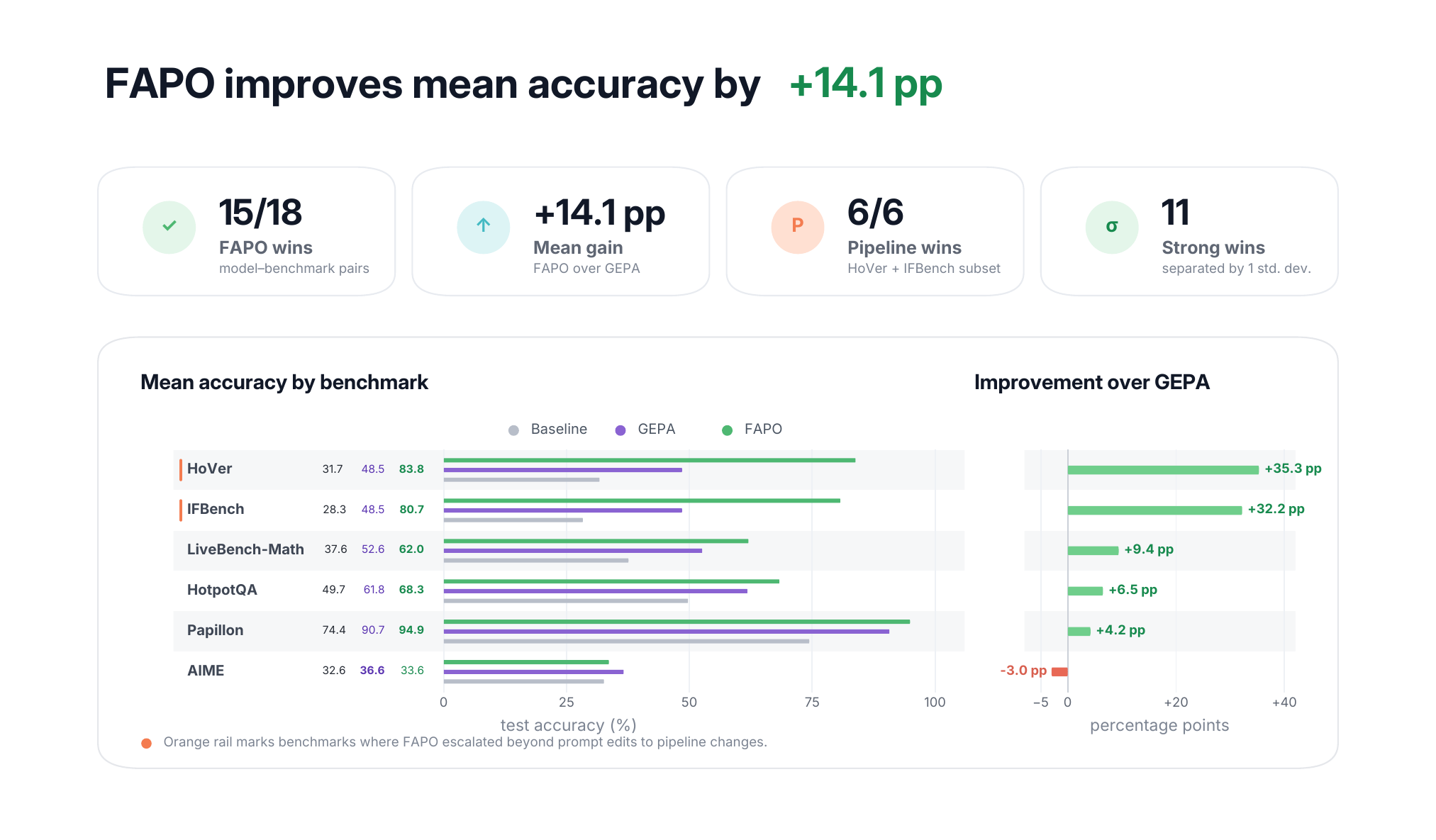}%
    }
    \caption{
Mean benchmark score across \GPTFourOneMini{}, \GPTFiveFourMini{}, and \GemmaThreeTwelveB{}. \textbf{\FAPO{} beats GEPA in 15 of 18 model--benchmark comparisons}, with a mean \FAPO{}--GEPA gain of \textbf{+14.1 percentage points}. The largest gains are on HoVer and IFBench, where \FAPO{} escalates from prompt edits to pipeline changes after prompt-level search exposes unresolved bottlenecks. AIME is the only benchmark where GEPA leads.
    }
    \label{fig:fapo-front-page-results}
\end{figure}
}

\section{Introduction}
\label{sec:intro}

Multi-step LLM pipelines are now common in security, enterprise analytics, and knowledge work.
They combine LLM-based calls with code-based steps to produce reliable workflows.
As workflow complexity and the number of LLM calls increase, traditional prompt optimization is not enough.
Failures can occur at any step and propagate through to downstream components.
Optimizing these systems requires more than single-turn prompt tuning.

Prompt-space search and optimization have already been extensively explored in the jailbreaking literature.
In red-teaming, the target is often adversarial and Best-of-\(N\): under a fixed query budget, generate or refine candidates until at least one prompt elicits a jailbreak.
Search strategies include simple parallel search~\citep{pair}, tree-based search~\citep{tap}, repeated sampling~\citep{bonjailbreak}, and heuristic search~\citep{advreasoning}, all aimed at finding at least one jailbreak prompt.
We use this closed-loop search pattern, but change the objective from finding one successful failure case to improving the mean score of one pipeline variant across \(N\) evaluation cases.
This objective shift makes attribution necessary: the optimizer must explain recurring failures rather than exploit a rare successful sample.

Existing tools leave a gap.
Evaluation suites such as HELM~\citep{helm}, BIG-bench~\citep{bigbench}, and AgentBench~\citep{agentbench} measure model capabilities.
However, they primarily evaluate model behavior over benchmark tasks rather than optimize the design of a fixed, inspectable pipeline.
Prompt-programming systems such as DSPy~\citep{dspy} optimize LLM-based modules; GEPA~\citep{gepa} optimizes prompts inside pipelines.
Neither is designed to inspect step-level failures and then change either prompts or pipeline structure inside a standard code workspace.

We present \textit{\textbf{F}ully \textbf{A}utomated \textbf{P}rompt \textbf{O}ptimization} (\FAPO{}).
\FAPO{} takes a well-structured problem statement, an evaluation criterion, and a task model, then searches for a higher-scoring pipeline for that task.
\FAPO{} has a reusable evaluation engine and isolated tenant workspaces.
\FAPO{} uses LangGraph~\citep{langgraph} to represent the pipeline as a stateful graph.
\ClaudeCode{}~\citep{claudecode} drives the optimization loop.
The agent analyzes failures, proposes variants, runs evaluations, and validates changes within tenant-defined guardrails.
The code is available at \url{https://github.com/cisco-foundation-ai/fully-automated-prompt-optimization}.

We evaluate \FAPO{} against GEPA across six benchmarks and three task models: \GPTFourOneMini{}~\citep{openai_gpt41}, \GPTFiveFourMini{}~\citep{openai_gpt5}, and \GemmaThreeTwelveB{}~\citep{gemma3}.
Both systems start from the same pipeline and baseline prompts.
\FAPO{} first attempts prompt-level optimization and escalates to structural changes only when attribution indicates that prompt edits are not enough to resolve the dominant bottleneck.
As shown in Figure~\ref{fig:fapo-front-page-results}, \FAPO{} wins 15 of 18 model--benchmark comparisons, with a mean gain of +14.1\,pp.
The largest improvements occur on HoVer~\citep{hover} and IFBench~\citep{ifbench}, where \FAPO{} extends retrieval chains or introduces deterministic constraint enforcement.
On prompt-only comparisons, \FAPO{} wins 9 of 12. We also consider CTIBench Root Cause Mapping~\citep{ctibench}, a security CVE-to-CWE classification task.
This experiment is constrained to prompt edits, following the Foundation-Sec evaluation protocol~\citep{foundationsec_instruct,foundationsec_reasoning}.
\FAPO{} improves performance for \GPTFive{}~\citep{openai_gpt5}, \FoundationSecEightBInstruct{}~\citep{foundationsec_instruct}, and \FoundationSecEightBReasoning{}~\citep{foundationsec_reasoning} on CTIBench-RCM.

The paper makes \underline{three contributions}:
\begin{itemize}
    \item \textbf{A \ClaudeCode{}-based pipeline optimization technique.}
    \FAPO{} starts with prompt edits and, when permitted, resorts to pipeline-structure edits only when recorded failure evidence indicates that prompt optimization is insufficient.
    \item \textbf{A reproducible workspace procedure for pipeline optimization.}
    \FAPO{} records final outputs, intermediate step outputs, configurations, and variant history.
    \item \textbf{Experiments demonstrating the technique's performance advantages.}
    The evaluation spans QA, fact verification, instruction following, math, and security classification.
\end{itemize}
We hope this work provides the community with accessible tooling for leveraging \ClaudeCode{}'s optimization capabilities across prompt and pipeline search.

\section{System Overview}
\label{sec:system}

\begin{figure}[t]
    \centering
    \includegraphics[width=0.98\textwidth]{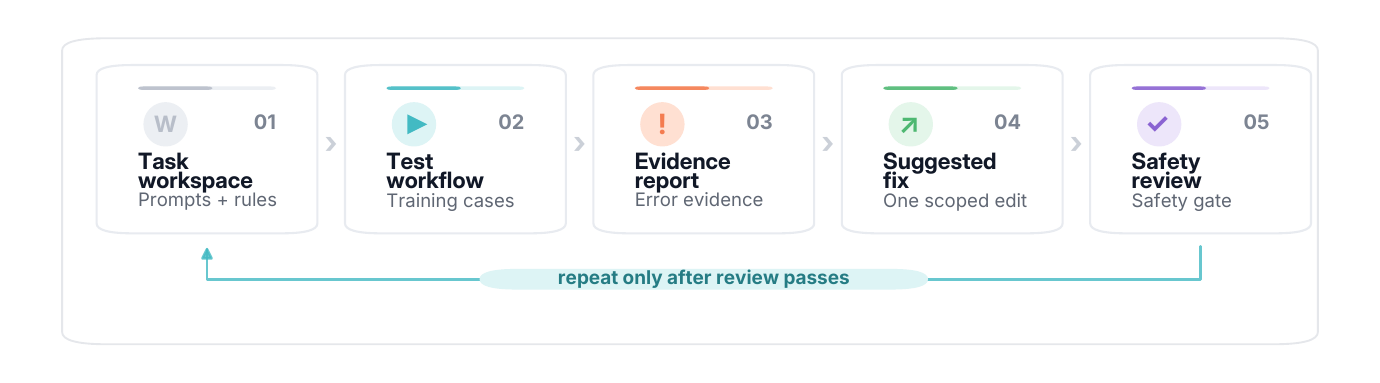}
    \caption{\FAPO{} as a reviewed improvement loop. The system tests the current workflow, records evidence from each step, proposes one allowed improvement, checks the proposal, and repeats only when the change passes review.}
    \label{fig:system-overview}
\end{figure}

\subsection{What \FAPO{} Does}
\label{sec:system-overview}

\FAPO{} treats an LLM pipeline as an inspectable workflow.
\FAPO{} records the inputs, outputs, and logs of each step in the pipeline.
The optimizer can then localize a failure to the prompt, an upstream evidence source, or the chain itself.

\begin{enumerate}
    \item \textbf{Start with one task workspace.} The workspace contains the task instructions, examples, scoring rule, current prompts, and allowed changes.
    \item \textbf{Run the current workflow.} A shared runner evaluates the pipeline on the training cases and records the final answer, score, and intermediate step outputs.
    \item \textbf{Find where mistakes begin.} The evidence report groups failures by likely cause, such as missing evidence, unsupported abstention, verbose answers, malformed output, or a weak final instruction.
    \item \textbf{Try one scoped change.} \ClaudeCode{} first proposes prompt edits. It later changes a parameter or adds a pipeline step only when prompt edits appear insufficient and the scope contract allows the change.
    \item \textbf{Review before rerunning.} A separate reviewer checks that the proposed change follows the task rules and does not leak data or change the scorer.
\end{enumerate}

\subsection{Tenant Organization}
\label{sec:tenant-organization}

\FAPO{} organizes optimization around a tenant, the unit used throughout the paper to represent a task with an evaluation criteria and workflows.
The core engine is the shared runtime under \code{src/hephaestus/}: it loads cases, renders prompts, calls provider adapters, runs LangGraph chains, validates scorer outputs, writes run artifacts, and supports failure attribution.
A tenant workspace under \code{tenants/<tenant\_id>/} contains the task-local material: chain code, prompt and chain variants, dataset conversion code and JSONL caches, scorers, eval configs, tests, storage configuration, operating documents, and optimization history.
The tenant playbook describes the tenant on a high-level, describes the layout of the tenant code and data, and specifies the constraints of the optimization.
The tenant playbook is treated as \textit{the most important} policy document during optimization and it can override \FAPO{} capabilities.
An eval config defines a reproducible chain configuration by specifying parameters as well as selecting variants (versions of prompts and chains that are generated during optimization).

We introduce this organizational structure to ensure reproducibility, isolation, and extensibility.
All variants and scores are recorded in the tenant-level directories for full visibility into prior runs and optimization attempts.
Tenants are isolated from each other to make sure that assumptions and invariants from one tenant do not affect any code or optimization in other tenants.
This model also allows each tenant to define their own pipelines, own scoring methods, own deployment methods, and more.\footnote{This tenant model is especially useful in corporate environments, where different business units, customers, security domains, or operational setups often impose idiosyncratic requirements and assumptions that are not academically clean, while still benefiting from a common evaluation runtime.}


\subsection{Design Principles}
\label{sec:system-design}

The architecture follows four principles.

\begin{itemize}
    \item \textbf{Separate the shared tester from the task.} The same runner can evaluate many tasks, while each task keeps its own prompts, examples, scoring rules, and change rules in its own workspace.
    \item \textbf{Ground decisions in recorded evidence.} \FAPO{} records intermediate steps so the optimizer can see whether a wrong answer came from retrieval, reasoning, formatting, or the final response step.
    \item \textbf{Prefer the smallest useful change.} \FAPO{} starts with prompt edits. It moves to settings or chain changes only when the recorded failures show that prompt optimization is no longer enough.
    \item \textbf{Keep optimization bounded.} The task workspace states what can and cannot be changed. The reviewer checks every proposed variant before it is evaluated.
\end{itemize}

In practice, a run follows the same pattern throughout the paper: evaluate the current pipeline, explain the failures, propose one allowed variant, review it, and keep it only if it improves validation performance.
Appendix~\ref{app:system-details} contains the implementation details: runtime configuration, chain state, scorer contracts, run artifacts, failure attribution, and tenant isolation.

\section{Claude-Driven Optimization}
\label{sec:optimization}

\FAPO{} uses \ClaudeCode{}~\citep{claudecode} as its orchestrator optimization layer, separate from the task model being evaluated.
It edits the workspace, runs evaluations, dispatches subagents, and records variants via custom skills and prompts in the \FAPO{} codebase.
This optimization mechanism can optimize pipelines that use a variety of closed or open-source task models.

\subsection{Implementation Components}
\label{sec:agents}

The optimization loop uses three core agents.
The \textbf{optimization agent} reads the tenant playbook and scope contract then drives the optimization loop.
The \textbf{step-attribution subagent} analyzes failures after each evaluation.
It uses rule-based checks and LLM analysis to classify failures as prompt-addressable or structural.
The \textbf{variant-reviewer subagent} checks each proposal for scope compliance, placeholder integrity, data leakage, and scorer compatibility.

\FAPO{} also provides \ClaudeCode{} with commands and repository instructions around the optimization loop (see Table~\ref{tab:claude-artifacts}).
These agents, commands, and skills provide guidance to \ClaudeCode{} on how to efficiently optimize without violating tenant constraints and guidance.

\begin{table}[t]
\centering
\scriptsize
\renewcommand{\arraystretch}{1.3}
\begin{tabular}{@{}llp{8.0cm}@{}}
\toprule
\textbf{Artifact} & \textbf{Type} & \textbf{Role} \\
\midrule
\code{optimization} & Agent & Orchestrates optimization: reads the tenant playbook, defines the scope contract, selects allowed optimization points, creates variants, runs evaluations, and records outcomes. \\
\code{step-attribution} & Agent & Analyzes eval results and failure modes using rule-based attribution with LLM-based inspection, clusters failures, and recommends the next optimization level. \\
\code{variant-reviewer} & Agent & Independently reviews proposed prompt or chain variants for scope compliance, placeholder integrity, leakage, scorer compatibility, state protocol compliance, and import safety. \\
\code{eval-runner} & Command & Runs a tenant eval config and returns score summaries plus the output directory. \\
\code{reset-tenant} & Command & Restores a tenant's working directory to baseline by removing non-baseline variants and optimization artifacts from the immediate directory while preserving those artifacts in git history. \\
\code{CLAUDE.md} & Repository instructions & Defines repository-wide Claude guidance: project purpose, eval workflow, code style, tenant data safety, and GitHub workflow conventions. \\
\bottomrule
\end{tabular}
\caption{\ClaudeCode{} artifacts provided by \FAPO{}. Core optimization uses the optimization, step-attribution, and variant-reviewer agents; the remaining commands and repository instructions support evaluation, recovery, and repo-wide guidance.}
\label{tab:claude-artifacts}
\end{table}

\subsection{The Optimization Loop}
\label{sec:opt-loop}

The optimizer first reads the tenant playbook.
It then writes a \textbf{scope contract}.
The contract states which optimization levels are allowed depending on the tenant instructions.
Currently three levels are possible: prompt text, chain parameters, or chain structure.

\begin{figure}[t]
  \centering
  \includegraphics[width=\linewidth]{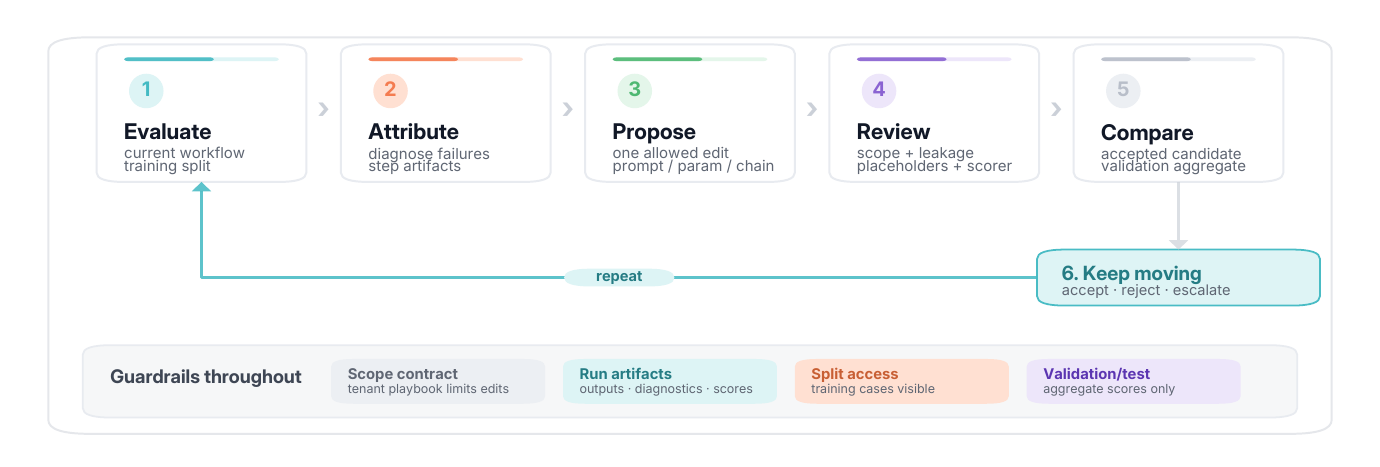}
  \caption{The \FAPO{} optimization loop. The optimizer evaluates the current variant, attributes failures using step-level artifacts, proposes one scoped change, sends it to the independent reviewer, compares accepted candidates on aggregate validation scores, and then either continues prompt optimization or escalates within the scope contract when prompt edits appear insufficient.}
  \label{fig:opt-loop}
\end{figure}

The loop then proceeds through six stages (Figure~\ref{fig:opt-loop}). First, \FAPO{} \textbf{evaluates} the current variant by running the pipeline on the training split and collecting final outputs together with intermediate-step evidence.
It then \textbf{attributes} failures by classifying them according to pipeline step and fix type.
Next, it \textbf{proposes} a scoped variant for the dominant failure cluster, and the reviewer \textbf{checks} the proposal for scope compliance, placeholder integrity, leakage, and scorer compatibility.
If the proposal passes review, \FAPO{} \textbf{evaluates} the proposed variant and \textbf{compares} it to the prior best variant.
Finally, it \textbf{iterates or escalates}: improved variants are kept, and when prompt-level search plateaus, the optimizer records the reason and explores a permitted non-prompt change only if failure analysis supports that escalation.

The optimizer chooses among allowed levels using the attribution report.
Prompt changes are the simplest option, so \FAPO{} tries them first.
It escalates to chain parameters or chain structure only when prompt-level optimization appears insufficient, the tenant scope contract permits those levels, and the attribution report identifies a bottleneck that prompts are unlikely to fix.
In the GEPA comparison, the non-CTIBench-RCM tenant scopes allowed chain-level variants; \FAPO{} still followed a prompt-first policy so that structural changes were considered only after prompt-level search exposed a structural bottleneck.

\subsection{Guardrails and Data Hygiene}
\label{sec:guardrails}

Automated optimization without constraints tends to overfit to examples.
\FAPO{} uses four guardrails:

\begin{itemize}
    \item \textbf{Split access controls}: The optimizer agent sees individual training cases. Validation and test expose aggregate scores only.
    \item \textbf{Scope constraints}: Tenant playbooks define allowed and forbidden changes. The optimizer and reviewer enforce them independently.
    \item \textbf{Iteration memory}: A structured log records variants, scores, and exhaustion reasons.
    \item \textbf{Variant immutability}: Every accepted or rejected attempt gets a new variant file.
\end{itemize}

\section{Evaluation}
\label{sec:eval}

We evaluate \FAPO{} against GEPA~\citep{gepa} across six benchmarks and three task models.

\subsection{Tasks}
\label{sec:tasks}

\paragraph{Multi-hop QA.}
We replicate the GEPA HotpotQA~\citep{hotpotqa} pipeline as a six-node LangGraph chain: two BM25 retrieval nodes ($k\!=\!7$) and four LLM nodes.
The optimization metric is exact match (EM), following GEPA's protocol; F1 is retained only as an auxiliary diagnostic.
Dataset splits follow GEPA: 150 development, 300 validation, and 300 test cases.
Baseline prompts use minimal DSPy-style instructions.

\paragraph{CTIBench Root Cause Mapping (RCM).}
CTIBench-RCM~\citep{ctibench} maps CVE descriptions to CWE IDs.
It is a 263-class security classification task.
We follow the Foundation-Sec setup~\citep{foundationsec_instruct,foundationsec_reasoning}: one classification node and exact-match scoring on extracted CWE IDs.
The dataset has 173 dev cases and 827 test cases.
Rare CWEs ($\leq$3 cases) appear only in test.
The optimizer is constrained to prompt edits.
We run on \GPTFive{}~\citep{openai_gpt5}, \FoundationSecEightBInstruct{}~\citep{foundationsec_instruct}, and \FoundationSecEightBReasoning{}~\citep{foundationsec_reasoning}.

\paragraph{HoVer.}
HoVer~\citep{hover} is a many-hop fact-verification task.
Each example gives a claim whose support or refutation depends on evidence spread across multiple Wikipedia articles.
The pipeline must retrieve the relevant evidence and classify the claim as supported or not supported.

\paragraph{Papillon.}
Papillon~\citep{papillon} evaluates privacy-conscious delegation.
The pipeline must preserve answer quality while limiting leakage of personally identifiable information when user requests are routed through a local-and-API model ensemble.
The task stresses prompt and pipeline choices that balance utility against privacy constraints.

\paragraph{IFBench.}
IFBench~\citep{ifbench} measures verifiable instruction following.
Each prompt contains explicit constraints on the required response.
The scorer checks whether the final answer satisfies those constraints, making format and constraint enforcement central failure modes.

\paragraph{LiveBench-Math.}
LiveBench-Math~\citep{livebench} evaluates mathematical reasoning on contamination-limited benchmark problems.
The pipeline must solve the problem and emit a final answer in a scoreable format.
Failures often come from either reasoning mistakes or final-answer extraction errors.

\paragraph{AIME.}
AIME~\citep{aime} uses competition-style mathematics problems with short exact answers.
The task emphasizes multi-step symbolic and numerical reasoning under strict answer formatting.

\subsection{Experimental Protocol}
\label{sec:protocol}

\paragraph{Models and optimization budget.}
For each benchmark in Table~\ref{tab:fapo-gepa}, FAPO starts optimization from the corresponding baseline GEPA pipeline.
Both systems start from identical baseline conditions: the same chain architecture, baseline prompts, task model, sampling parameters, metric, and splits.
After the baseline run, the optimization scopes differ.
GEPA searches instruction strings in the reproduced DSPy program, while the \FAPO{} scope contracts for these non-CTIBench-RCM tasks allowed prompt, chain-parameter, and chain-architecture variants under a prompt-first escalation policy.
Sampling uses temperature~1.0, top-\(p\)~0.95, and a nominal 16{,}000-token generation limit. Three task models are evaluated: \GPTFourOneMini{}~\citep{openai_gpt41}, \GPTFiveFourMini{}~\citep{openai_gpt5}, and \GemmaThreeTwelveB{}~\citep{gemma3}.
For \GPTFiveFourMini{}, which the provider offers as a reasoning model, the token limit corresponds to a shared \texttt{max\_completion\_tokens} budget covering both hidden reasoning and visible output; for the remaining two, it applies to visible output only, as these models do not perform reasoning.
The \FAPO{} budget is limited to 50 variants or 10 optimization rounds per trial, whichever comes first.
No early stopping is applied within these bounds.

\paragraph{GEPA reproduction.}
GEPA optimizes the instruction string inside a fixed DSPy chain-of-thought program using MIPROv2-Heavy evolutionary search.
We use the authors' code as-is, except for the reflector model, which we replace with \ClaudeOpusFourSix{} through Amazon Bedrock using provider-default settings.
This gives GEPA a strong contemporary reflector, but it does not make the two systems identical: GEPA remains a fixed-program prompt optimizer, whereas \FAPO{} is an agentic workspace optimizer with a different search space.

\paragraph{Trial protocol.}
Each cell in Table~\ref{tab:fapo-gepa} reports mean test score $\pm$ standard deviation over three trials.
The reported score is the test score of the best validation-selected variant from that trial.
The Baseline column reports refreshed pristine variant-001 test scores; validation baselines were refreshed from the same pristine setup before validation selection and trajectory plotting.
For this comparison, \FAPO{} starts at prompt level.
For the non-CTIBench-RCM tasks, the scope contract permits escalation to chain-parameter or chain-architecture changes only when prompt optimization appears insufficient and attribution finds a structural bottleneck.
CTIBench-RCM remains prompt-only.

\begin{table}[tbh]
\centering
\small
    \begin{tabular}{@{}ll@{\hskip 24pt}c@{\hskip 12pt}c@{\hskip 12pt}c@{\hskip 18pt}r@{}}
        \toprule
        \textbf{Benchmark} & \textbf{Model} & \textbf{Baseline} & \textbf{GEPA} & \textbf{\FAPO{}} & \textbf{$\Delta$} \\
        \midrule
        \multirow{3}{*}[-2pt]{HotpotQA} & \GPTFourOneMini{} & $37.11 \pm 1.07$ & $67.56 \pm 3.83$ & \highlight[customblue]{$\mathbf{72.67} \pm 3.20$} & $+5.11$ \\
         & \GPTFiveFourMini{} & $55.56 \pm 3.20$ & $56.22 \pm 3.47$ & \highlight{$\mathbf{69.56} \pm 1.50$} & $+13.34$ \\
         & \GemmaThreeTwelveB{} & $56.56 \pm 1.92$ & $61.66 \pm 0.58$ & \highlight{$\mathbf{62.78} \pm 0.19$} & $+1.12$ \\
        \addlinespace[3pt]\midrule\addlinespace[3pt]
        \multirow{3}{*}[-2pt]{HoVer\textsuperscript{\textdaggerdbl}} & \GPTFourOneMini{} & $43.89 \pm 14.92$ & $59.67 \pm 1.21$ & \highlight{$\mathbf{84.44} \pm 17.27$} & $+24.78$ \\
         & \GPTFiveFourMini{} & $26.33 \pm 1.00$ & $31.78 \pm 2.14$ & \highlight{$\mathbf{80.33} \pm 9.60$} & $+48.56$ \\
         & \GemmaThreeTwelveB{} & $24.89 \pm 13.66$ & $54.00 \pm 2.02$ & \highlight{$\mathbf{86.67} \pm 5.61$} & $+32.67$ \\
        \addlinespace[3pt]\midrule\addlinespace[3pt]
        \multirow{3}{*}[-2pt]{IFBench\textsuperscript{\textdaggerdbl}} & \GPTFourOneMini{} & $25.74 \pm 0.71$ & $54.76 \pm 1.11$ & \highlight{$\mathbf{93.71} \pm 9.75$} & $+38.95$ \\
         & \GPTFiveFourMini{} & $24.49 \pm 1.23$ & $48.36 \pm 0.39$ & \highlight{$\mathbf{86.05} \pm 10.79$} & $+37.70$ \\
         & \GemmaThreeTwelveB{} & $34.69 \pm 0.74$ & $42.46 \pm 2.31$ & \highlight[customblue]{$\mathbf{62.30} \pm 28.75$} & $+19.84$ \\
        \addlinespace[3pt]\midrule\addlinespace[3pt]
        \multirow{3}{*}[-2pt]{LiveBench-Math} & \GPTFourOneMini{} & $50.25 \pm 1.95$ & $61.78 \pm 1.57$ & \highlight{$\mathbf{73.56} \pm 1.02$} & $+11.78$ \\
         & \GPTFiveFourMini{} & $31.76 \pm 1.51$ & $57.26 \pm 0.28$ & \highlight{$\mathbf{67.00} \pm 6.35$} & $+9.73$ \\
         & \GemmaThreeTwelveB{} & $30.80 \pm 1.63$ & $38.66 \pm 2.69$ & \highlight{$\mathbf{45.30} \pm 2.29$} & $+6.64$ \\
        \addlinespace[3pt]\midrule\addlinespace[3pt]
        \multirow{3}{*}[-2pt]{AIME} & \GPTFourOneMini{} & $51.11 \pm 5.09$ & \highlight[customblue]{$\mathbf{52.67} \pm 7.43$} & $48.89 \pm 0.38$ & $-3.78$ \\
         & \GPTFiveFourMini{} & $30.00 \pm 6.67$ & \highlight[customblue]{$\mathbf{38.44} \pm 1.93$} & $33.78 \pm 12.11$ & $-4.67$ \\
         & \GemmaThreeTwelveB{} & $16.67 \pm 3.33$ & \highlight[customblue]{$\mathbf{18.67} \pm 4.81$} & $18.22 \pm 1.68$ & $-0.44$ \\
        \addlinespace[3pt]\midrule\addlinespace[3pt]
        \multirow{3}{*}[-2pt]{Papillon} & \GPTFourOneMini{} & $68.39 \pm 0.80$ & $90.72 \pm 3.61$ & \highlight[customblue]{$\mathbf{94.29} \pm 0.96$} & $+3.57$ \\
         & \GPTFiveFourMini{} & $87.79 \pm 1.00$ & $88.82 \pm 2.20$ & \highlight{$\mathbf{95.07} \pm 0.99$} & $+6.26$ \\
         & \GemmaThreeTwelveB{} & $67.15 \pm 1.07$ & $92.65 \pm 2.25$ & \highlight[customblue]{$\mathbf{95.47} \pm 1.09$} & $+2.81$ \\
        \bottomrule
    \end{tabular}
\caption{Comparison of \FAPO{} and GEPA across six benchmarks and three task models. Scores report test benchmark metric (\%) averaged over three trials $\pm$ trial standard deviation; HotpotQA uses EM. Boldface marks the higher mean between GEPA and \FAPO{}. Orange shading marks wins with non-overlapping mean $\pm$ trial-standard-deviation ranges; blue shading marks wins with overlapping ranges. $\Delta$ = \FAPO{} $-$ GEPA, computed from unrounded trial means. \FAPO{} wins 15 of 18 comparisons; 11 wins have non-overlapping mean $\pm$ trial-standard-deviation ranges. {\textsuperscript{\textdaggerdbl}\,\FAPO{} used permitted pipeline optimization after prompt-level search indicated that prompt edits were insufficient. We used the pipeline shape from GEPA as the baseline pipeline for each non-CTIBench-RCM task.}}
\label{tab:fapo-gepa}
\end{table}

\subsection{Results}
\label{sec:results}

Table~\ref{tab:fapo-gepa} reports the results.
FAPO-optimized pipelines typically outperform GEPA-optimized chains, except for the AIME benchmark; see the subsequent discussion. On two benchmarks -- HoVer and IFBench -- FAPO had to escalate to pipeline optimization. On HoVer, attribution identified insufficient retrieval coverage. \FAPO{} extended the baseline 3-hop retrieval chain to 4--5 hops, with multi-query BM25 search and entity-aware rescue.
On IFBench, attribution identified format failures. \FAPO{} added deterministic post-processing nodes that enforce instruction constraints. These changes produce gains of +24.78 to +48.56\,pp on HoVer and +19.84 to +38.95\,pp on IFBench.

On the remaining benchmarks, optimization stayed at the prompt level. \FAPO{} wins 9 of 12 prompt-only comparisons, six of which have non-overlapping mean $\pm$ trial-standard-deviation ranges, suggesting statistically significant improvements. AIME is the only benchmark where GEPA leads \FAPO{} across all three model comparisons; relative to the baselines, \FAPO{} yields mixed results ($-2.22$, $+3.78$, and $+1.55$\,pp across the three task models) that fall within the noise range, so we treat the AIME result as inconclusive rather than a consistent prompt-optimization gain. We speculate that the inconsistent AIME results may stem from overfitting to small sample sizes relative to the problem space.

\subsection{Case Studies}
\label{sec:case-studies}

\begin{figure}[t]
    \centering
    \includegraphics[width=\linewidth]{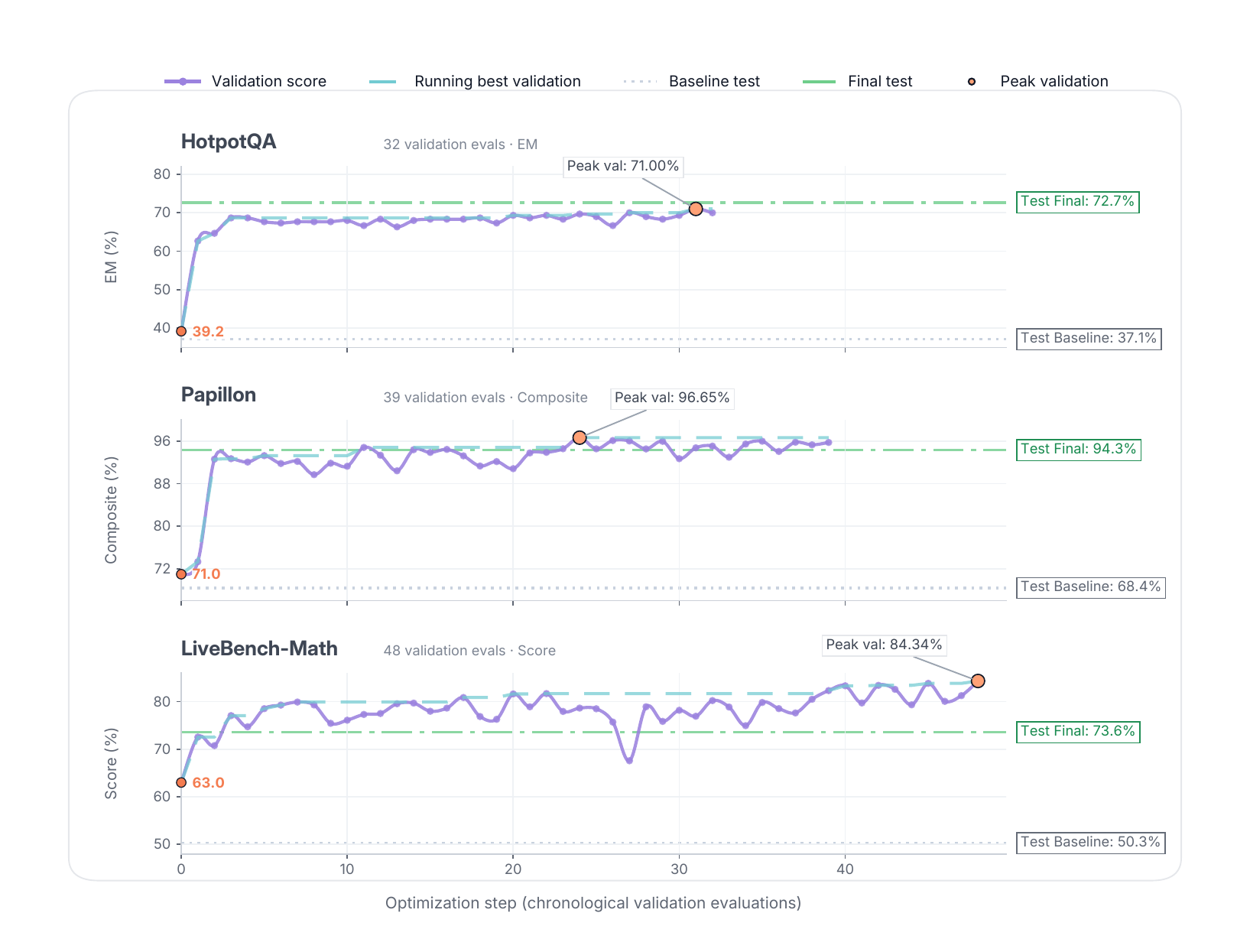}
    \caption{
        Evolution of \FAPO{} variants over time across HotpotQA, Papillon, and LiveBench-Math, using \GPTFourOneMini{} as the evaluated task model.
        Solid lines show validation performance on the benchmark optimization metric (EM for HotpotQA); dashed lines show the running-best validation score.
        Horizontal reference lines mark the baseline test score and final \FAPO{} test score measured with the same benchmark metric; orange markers identify peak validation scores.
    }
    \label{fig:fapo-progression}
\end{figure}

\paragraph{HotpotQA.}
Figure~\ref{fig:fapo-progression} shows \GPTFourOneMini{} validation trajectories for HotpotQA, Papillon, and LiveBench-Math.
The HotpotQA trajectory reports exact match (EM), the optimization metric used for GEPA compatibility.
For HotpotQA, the refreshed pristine baseline scored $39.22 \pm 1.17$\% validation EM and $37.11 \pm 1.07$\% test EM.
The attribution subagent identified three failure categories on the dev set: near-miss (verbose answers, 13 cases), abstention (model declined to answer, 8 cases), and wrong-answer (17 cases).
Variant-002 addressed near-miss failures with answer brevity constraints, raising validation EM to 65.7\%; variant-003 addressed abstention failures with a must-always-answer rule, raising validation EM to 70.3\%.
After two iterations the attribution system flagged remaining failures as retrieval-limited (structural), indicating that further prompt-only iteration was unlikely to help.
The validation-selected HotpotQA variant in Table~\ref{tab:fapo-gepa} remained prompt-only.

\paragraph{CTIBench-RCM.}
Table~\ref{tab:rcm} shows CTIBench-RCM results.
Each model was optimized independently under prompt-only scope.
The final histories contain 31, 30, and 27 tested variants, for 88 variants across the three models. The best strategy differs by model. \GPTFive{} improved from 72.1\% to 76.1\% test accuracy after adding NVD convention rules for common CWE confusions. \FoundationSecEightBInstruct{} improved from 63.9\% to 71.0\% with a shorter prompt that mentions NVD mapping conventions. \FoundationSecEightBReasoning{} improved from 71.0\% to 73.0\% with the phrase ``standard NVD abstraction level.'' Most remaining errors come from ambiguous CWE labels, especially CWE-787 versus CWE-121/122 in buffer overflow descriptions.

\begin{table}[!t]
\centering
\small
    \begin{tabular}{@{}lccccc@{}}
    \toprule
        \textbf{Model} & \textbf{Variants} & \textbf{Dev (base)} & \textbf{Dev (best)} & \textbf{Test (base)} & \textbf{Test (best)}  \\
        \midrule
        \GPTFive{} & 31 & 78.6 & 85.6 & 72.1 & \textbf{76.1} \\
        \FoundationSecEightBInstructShort{} & 30 & 76.3 & 80.4 & 63.9 & \textbf{71.0} \\
        \FoundationSecEightBReasoningShort{} & 27 & 80.4 & 83.2 & 71.0 & \textbf{73.0} \\
        \bottomrule
        \end{tabular}
\caption{CTIBench-RCM optimization. All scores are exact-match accuracy (\%). Optimizer scope is prompt-only.}
\label{tab:rcm}
\end{table}

\subsection{Discussion}
\label{sec:discussion}

\paragraph{Experimental design.}
The comparison in Table~\ref{tab:fapo-gepa} gives \FAPO{} a broader optimization scope than GEPA's prompt optimizer, which searches over instruction strings within a fixed DSPy program. For the GEPA-comparison tasks other than CTIBench-RCM, \FAPO{} was permitted to modify chain parameters and structure (by its nature), starting from the same baseline pipeline but only after first attempting prompt optimization. Rows where \FAPO{} modified the chain architecture are marked in the table. These results suggest that pipeline modifications can yield improvements beyond the reach of prompt-only search. The prompt-only subset still favors \FAPO{} -- 9 of 12 wins, 6 with non-overlapping mean $\pm$ trial-standard-deviation ranges -- an advantage we attribute to the deep, iterative reasoning of the Claude Code orchestrator.

\paragraph{Trial variance.}
\FAPO{} has higher run-to-run variation when prompt-first search is allowed to escalate to pipeline changes.
This reflects path dependence in the optimization trajectory.
In some trials, \FAPO{} escalates from prompt edits to architecture changes after identifying a structural bottleneck.
In others, it remains at the prompt level and behaves more like a prompt optimizer.
Thus, the larger standard deviations mainly reflect whether the optimization trajectory discovers a structural intervention, rather than a smooth spread of outcomes around one typical variant.

\paragraph{Baseline model asymmetry.}
\GPTFourOneMini{} outperforms \GPTFiveFourMini{} on four of the six baseline test benchmarks.
Both models were given an identical 16{,}000-token generation budget, but the provider accounts for that budget differently: for \GPTFourOneMini{} it caps visible output only, whereas for \GPTFiveFourMini{} it is a shared \texttt{max\_completion\_tokens} budget covering hidden reasoning and visible output.
Reconstructed output-token counts confirm the asymmetry.
\GPTFourOneMini{} reaches the 16{,}000-token ceiling on long-derivation tasks such as AIME and LiveBench-Math and occasionally truncates, whereas \GPTFiveFourMini{}'s visible output never exceeds 4{,}922 tokens across 3{,}813 cases.
On the same tasks, \GPTFiveFourMini{} frequently emits very short or malformed final answers---LiveBench-Math has median visible output length 78 tokens---that fail the strict answer-extraction scorers: the \texttt{\textbackslash boxed\{\}} or trailing-integer parser for AIME, and exact match for LiveBench-Math.
This helps explain its lower baseline scores despite being the newer reasoning-capable model.
Tenant-level optimization logs corroborate the shared-budget effect: \GPTFiveFourMini{} required raising the budget to 24k--28k and enabling high reasoning effort to become competitive, with explicit over-/under-thinking trade-offs at 32k.
This asymmetry does not affect the controlled nature of the comparison: baseline prompts and evaluation settings are held fixed across models, and \FAPO{} optimizes each model independently.

\paragraph{GEPA reproduction fidelity.}
Our reproduced GEPA scores differ from the published results by $-3.78$ to $+7.97$\,pp.
GEPA and \FAPO{} differ in optimizer design and allowable search space, so the comparison should be read as a reproduced benchmark comparison rather than an exact fairness match.
We use \ClaudeOpusFourSix{} through Bedrock with default settings as GEPA's reflector, replacing the original reflector model; this may strengthen GEPA on HoVer and IFBench relative to the reported scores.
The original paper reports single-trial scores, whereas we report three-trial means and standard deviations.
All other parameters match the released GEPA codebase.\footnote{\url{https://github.com/gepa-ai/gepa}}

\section{Related Work}
\label{sec:related}

\paragraph{Pipeline and prompt optimization.}
Pipeline optimization improves multi-step LLM systems at granularities ranging from prompt text to module composition and chain topology.
GEPA~\citep{gepa} uses evolutionary search to optimize prompts for multi-step reasoning pipelines.
DSPy~\citep{dspy} compiles declarative LLM programs into optimized pipelines; MIPRO~\citep{mipro} extends DSPy with joint optimization of instructions and demonstrations for multi-stage programs.
APE~\citep{ape} frames instruction generation as black-box optimization, using an LLM to propose and score candidate prompts.
OPRO~\citep{opro} embeds an ``optimization trajectory'' of past candidates and scores directly in the prompt, using the LLM itself as the optimizer.
EvoPrompt~\citep{evoprompt} and PromptBreeder~\citep{promptbreeder} apply evolutionary algorithms---with LLM-assisted mutation operators---to maintain populations of candidate prompts.
TextGrad~\citep{textgrad} treats textual feedback as a gradient-like signal over a computation graph of LLM calls, optimizing prompts as differentiable variables.
\FAPO{} builds on GEPA's evaluation setup but changes the optimizer: attribution-driven \ClaudeCode{} agents analyze failures, first propose prompt variants, and move to chain parameters or chain structure only when prompt optimization appears insufficient.
\FAPO{} is distinct from this line of work because it combines pipeline-aware step-level attribution, prompt-first multi-level optimization that escalates from prompt text through chain parameters to structural changes only when evidence supports it, scope-constrained guardrails, and multi-tenant isolation.

\paragraph{Automated research agents.}
\citet{karpathy_autoresearch} presents a minimalist ``autoresearch'' loop in which an LLM agent edits a single \code{train.py} file for a small LLM training setup, runs fixed five-minute single-GPU experiments, and keeps or reverts code changes according to validation bits per byte.
Autoresearch optimizes model-training code and hyperparameters under one scalar training metric.
\FAPO{} shares the idea of agent-driven closed-loop experimentation, but targets optimization on discrete landscapes with non-differentiable objective functions.

\paragraph{From jailbreaking to prompt optimization.}
Automated jailbreaking treats prompts as a discrete action space, uses a verifier or score as feedback, and spends test-time compute to search that space.
In Best-of-\(N\) red-teaming, success is existential: among \(N\) sampled or optimized candidates, the attack succeeds if any candidate jailbreaks the target~\citep{bonjailbreak}.
Prior jailbreaking work explores the same adversarial optimization view at richer search levels:
TAP searches a tree of attacker-proposed prompts~\citep{tap};
capability-scaling work studies how attacker and target capability affect jailbreak success~\citep{capscaling};
and adversarial reasoning frames jailbreaking as test-time optimization over reasoning strings~\citep{advreasoning}.

The technical lineage begins with universal adversarial triggers~\citep{triggers}, which used gradient-guided token search to find input-agnostic sequences that transfer across examples and models.
AutoPrompt~\citep{autoprompt} applied the same gradient-guided discrete search to \emph{improve task performance}.
GCG~\citep{gcg} then adapted token-level optimization to aligned chat models, producing universal adversarial suffixes that transfer to black-box targets---and explicitly framing the method as ``automated prompt generation.''
At the semantic level, PAIR~\citep{pair} used an attacker LLM to iteratively refine jailbreak prompts with only black-box access;
TAP~\citep{tap} scaled this into a tree search with pruning, reporting high success rates on frontier models with reduced query budgets.
AutoDAN~\citep{autodan} applied genetic algorithms with a stealthiness constraint---isomorphic to constraint satisfaction in benign prompt optimization.
Recent systems make the dual-use connection explicit:
EvoX~\citep{evox} meta-evolves both candidate prompts and the search strategies that generate them;
AdaEvolve~\citep{adaevolve} adds hierarchical adaptive scheduling to LLM-driven evolutionary search;
Claudini~\citep{claudini} uses \ClaudeCode{} agents to iteratively discover white-box adversarial attacks that recombine GCG variants---the same evaluate--analyze--propose--iterate loop that \FAPO{} applies to constructive pipeline improvement.

\FAPO{} is a constructive continuation of this search pattern.
It keeps the evaluate--analyze--propose--iterate loop but changes the target to aggregate validation performance for one deployable pipeline variant.
Instead of maximizing the probability of a rare successful failure, it improves mean behavior across examples while preserving task constraints.
\FAPO{} focuses on stable constructive improvement under multi-step pipeline attribution rather than adversarial single-success search.

\section{Conclusion}
\label{sec:conclusion}

Multi-step LLM pipelines fail through interactions among retrieval, reasoning, formatting, and control flow.
Improving them requires more than tuning one prompt in isolation.
We present \FAPO{}, a \ClaudeCode{}-based framework that turns those failures into a reproducible optimization loop: evaluate the pipeline, inspect intermediate steps, diagnose the bottleneck, propose a scoped change, and validate the resulting variant.
\FAPO{} starts with prompt edits and escalates to structure only when attribution indicates that the pipeline itself is limiting performance.
This procedure outperforms GEPA in 15 of 18 model--benchmark comparisons using both prompt and chain-level optimizations and improves three security models' performance on CTIBench-RCM.
These results show that pipeline-aware, evidence-grounded optimization can serve both general-purpose and security-focused tasks, and position \FAPO{} as a practical path toward more reliable multi-step LLM systems.

{\footnotesize
\setlength{\bibsep}{2pt}
\bibliographystyle{plainnat}
\bibliography{references}

@inproceedings{
    gepa,
    title={{GEPA}: Reflective Prompt Evolution Can Outperform Reinforcement Learning},
    author={Lakshya A Agrawal and Shangyin Tan and Dilara Soylu and Noah Ziems and Rishi Khare and Krista Opsahl-Ong and Arnav Singhvi and Herumb Shandilya and Michael J Ryan and Meng Jiang and Christopher Potts and Koushik Sen and Alex Dimakis and Ion Stoica and Dan Klein and Matei Zaharia and Omar Khattab},
    booktitle={The Fourteenth International Conference on Learning Representations},
    year={2026},
    url={https://openreview.net/forum?id=RQm2KQTM5r}
}

@inproceedings{
    dspy,
    title={{DSP}y: Compiling Declarative Language Model Calls into State-of-the-Art Pipelines},
    author={Omar Khattab and Arnav Singhvi and Paridhi Maheshwari and Zhiyuan Zhang and Keshav Santhanam and Sri Vardhamanan A and Saiful Haq and Ashutosh Sharma and Thomas T. Joshi and Hanna Moazam and Heather Miller and Matei Zaharia and Christopher Potts},
    booktitle={The Twelfth International Conference on Learning Representations},
    year={2024},
    url={https://openreview.net/forum?id=sY5N0zY5Od}
}

@inproceedings{hotpotqa,
    title = "{H}otpot{QA}: A Dataset for Diverse, Explainable Multi-hop Question Answering",
    author = "Yang, Zhilin  and
      Qi, Peng  and
      Zhang, Saizheng  and
      Bengio, Yoshua  and
      Cohen, William  and
      Salakhutdinov, Ruslan  and
      Manning, Christopher D.",
    editor = "Riloff, Ellen  and
      Chiang, David  and
      Hockenmaier, Julia  and
      Tsujii, Jun{'}ichi",
    booktitle = "Proceedings of the 2018 Conference on Empirical Methods in Natural Language Processing",
    month = oct # "-" # nov,
    year = "2018",
    address = "Brussels, Belgium",
    publisher = "Association for Computational Linguistics",
    url = "https://aclanthology.org/D18-1259/",
    doi = "10.18653/v1/D18-1259",
    pages = "2369--2380"
}

@misc{langgraph,
  title={{LangGraph}: Building Stateful, Multi-Agent Applications with {LLM}s},
  author={{LangChain}},
  year={2024},
  note={\url{https://github.com/langchain-ai/langgraph}}
}

@article{
bigbench,
title={Beyond the Imitation Game: Quantifying and extrapolating the capabilities of language models},
author={Aarohi Srivastava and Abhinav Rastogi and Abhishek Rao and Abu Awal Md Shoeb and Abubakar Abid and Adam Fisch and Adam R. Brown and Adam Santoro and Aditya Gupta and Adri{\`a} Garriga-Alonso and Agnieszka Kluska and Aitor Lewkowycz and Akshat Agarwal and Alethea Power and Alex Ray and Alex Warstadt and Alexander W. Kocurek and Ali Safaya and Ali Tazarv and Alice Xiang and Alicia Parrish and Allen Nie and Aman Hussain and Amanda Askell and Amanda Dsouza and Ambrose Slone and Ameet Rahane and Anantharaman S. Iyer and Anders Johan Andreassen and Andrea Madotto and Andrea Santilli and Andreas Stuhlm{\"u}ller and Andrew M. Dai and Andrew La and Andrew Kyle Lampinen and Andy Zou and Angela Jiang and Angelica Chen and Anh Vuong and Animesh Gupta and Anna Gottardi and Antonio Norelli and Anu Venkatesh and Arash Gholamidavoodi and Arfa Tabassum and Arul Menezes and Arun Kirubarajan and Asher Mullokandov and Ashish Sabharwal and Austin Herrick and Avia Efrat and Aykut Erdem and Ayla Karaka{\c{s}} and B. Ryan Roberts and Bao Sheng Loe and Barret Zoph and Bart{\l}omiej Bojanowski and Batuhan {\"O}zyurt and Behnam Hedayatnia and Behnam Neyshabur and Benjamin Inden and Benno Stein and Berk Ekmekci and Bill Yuchen Lin and Blake Howald and Bryan Orinion and Cameron Diao and Cameron Dour and Catherine Stinson and Cedrick Argueta and Cesar Ferri and Chandan Singh and Charles Rathkopf and Chenlin Meng and Chitta Baral and Chiyu Wu and Chris Callison-Burch and Christopher Waites and Christian Voigt and Christopher D Manning and Christopher Potts and Cindy Ramirez and Clara E. Rivera and Clemencia Siro and Colin Raffel and Courtney Ashcraft and Cristina Garbacea and Damien Sileo and Dan Garrette and Dan Hendrycks and Dan Kilman and Dan Roth and C. Daniel Freeman and Daniel Khashabi and Daniel Levy and Daniel Mosegu{\'\i} Gonz{\'a}lez and Danielle Perszyk and Danny Hernandez and Danqi Chen and Daphne Ippolito and Dar Gilboa and David Dohan and David Drakard and David Jurgens and Debajyoti Datta and Deep Ganguli and Denis Emelin and Denis Kleyko and Deniz Yuret and Derek Chen and Derek Tam and Dieuwke Hupkes and Diganta Misra and Dilyar Buzan and Dimitri Coelho Mollo and Diyi Yang and Dong-Ho Lee and Dylan Schrader and Ekaterina Shutova and Ekin Dogus Cubuk and Elad Segal and Eleanor Hagerman and Elizabeth Barnes and Elizabeth Donoway and Ellie Pavlick and Emanuele Rodol{\`a} and Emma Lam and Eric Chu and Eric Tang and Erkut Erdem and Ernie Chang and Ethan A Chi and Ethan Dyer and Ethan Jerzak and Ethan Kim and Eunice Engefu Manyasi and Evgenii Zheltonozhskii and Fanyue Xia and Fatemeh Siar and Fernando Mart{\'\i}nez-Plumed and Francesca Happ{\'e} and Francois Chollet and Frieda Rong and Gaurav Mishra and Genta Indra Winata and Gerard de Melo and Germ{\`a}n Kruszewski and Giambattista Parascandolo and Giorgio Mariani and Gloria Xinyue Wang and Gonzalo Jaimovitch-Lopez and Gregor Betz and Guy Gur-Ari and Hana Galijasevic and Hannah Kim and Hannah Rashkin and Hannaneh Hajishirzi and Harsh Mehta and Hayden Bogar and Henry Francis Anthony Shevlin and Hinrich Schuetze and Hiromu Yakura and Hongming Zhang and Hugh Mee Wong and Ian Ng and Isaac Noble and Jaap Jumelet and Jack Geissinger and Jackson Kernion and Jacob Hilton and Jaehoon Lee and Jaime Fern{\'a}ndez Fisac and James B Simon and James Koppel and James Zheng and James Zou and Jan Kocon and Jana Thompson and Janelle Wingfield and Jared Kaplan and Jarema Radom and Jascha Sohl-Dickstein and Jason Phang and Jason Wei and Jason Yosinski and Jekaterina Novikova and Jelle Bosscher and Jennifer Marsh and Jeremy Kim and Jeroen Taal and Jesse Engel and Jesujoba Alabi and Jiacheng Xu and Jiaming Song and Jillian Tang and Joan Waweru and John Burden and John Miller and John U. Balis and Jonathan Batchelder and Jonathan Berant and J{\"o}rg Frohberg and Jos Rozen and Jose Hernandez-Orallo and Joseph Boudeman and Joseph Guerr and Joseph Jones and Joshua B. Tenenbaum and Joshua S. Rule and Joyce Chua and Kamil Kanclerz and Karen Livescu and Karl Krauth and Karthik Gopalakrishnan and Katerina Ignatyeva and Katja Markert and Kaustubh Dhole and Kevin Gimpel and Kevin Omondi and Kory Wallace Mathewson and Kristen Chiafullo and Ksenia Shkaruta and Kumar Shridhar and Kyle McDonell and Kyle Richardson and Laria Reynolds and Leo Gao and Li Zhang and Liam Dugan and Lianhui Qin and Lidia Contreras-Ochando and Louis-Philippe Morency and Luca Moschella and Lucas Lam and Lucy Noble and Ludwig Schmidt and Luheng He and Luis Oliveros-Col{\'o}n and Luke Metz and L{\"u}tfi Kerem Senel and Maarten Bosma and Maarten Sap and Maartje Ter Hoeve and Maheen Farooqi and Manaal Faruqui and Mantas Mazeika and Marco Baturan and Marco Marelli and Marco Maru and Maria Jose Ramirez-Quintana and Marie Tolkiehn and Mario Giulianelli and Martha Lewis and Martin Potthast and Matthew L Leavitt and Matthias Hagen and M{\'a}ty{\'a}s Schubert and Medina Orduna Baitemirova and Melody Arnaud and Melvin McElrath and Michael Andrew Yee and Michael Cohen and Michael Gu and Michael Ivanitskiy and Michael Starritt and Michael Strube and Micha{\l} Sw{\k{e}}drowski and Michele Bevilacqua and Michihiro Yasunaga and Mihir Kale and Mike Cain and Mimee Xu and Mirac Suzgun and Mitch Walker and Mo Tiwari and Mohit Bansal and Moin Aminnaseri and Mor Geva and Mozhdeh Gheini and Mukund Varma T and Nanyun Peng and Nathan Andrew Chi and Nayeon Lee and Neta Gur-Ari Krakover and Nicholas Cameron and Nicholas Roberts and Nick Doiron and Nicole Martinez and Nikita Nangia and Niklas Deckers and Niklas Muennighoff and Nitish Shirish Keskar and Niveditha S. Iyer and Noah Constant and Noah Fiedel and Nuan Wen and Oliver Zhang and Omar Agha and Omar Elbaghdadi and Omer Levy and Owain Evans and Pablo Antonio Moreno Casares and Parth Doshi and Pascale Fung and Paul Pu Liang and Paul Vicol and Pegah Alipoormolabashi and Peiyuan Liao and Percy Liang and Peter W Chang and Peter Eckersley and Phu Mon Htut and Pinyu Hwang and Piotr Mi{\l}kowski and Piyush Patil and Pouya Pezeshkpour and Priti Oli and Qiaozhu Mei and Qing Lyu and Qinlang Chen and Rabin Banjade and Rachel Etta Rudolph and Raefer Gabriel and Rahel Habacker and Ramon Risco and Rapha{\"e}l Milli{\`e}re and Rhythm Garg and Richard Barnes and Rif A. Saurous and Riku Arakawa and Robbe Raymaekers and Robert Frank and Rohan Sikand and Roman Novak and Roman Sitelew and Ronan Le Bras and Rosanne Liu and Rowan Jacobs and Rui Zhang and Russ Salakhutdinov and Ryan Andrew Chi and Seungjae Ryan Lee and Ryan Stovall and Ryan Teehan and Rylan Yang and Sahib Singh and Saif M. Mohammad and Sajant Anand and Sam Dillavou and Sam Shleifer and Sam Wiseman and Samuel Gruetter and Samuel R. Bowman and Samuel Stern Schoenholz and Sanghyun Han and Sanjeev Kwatra and Sarah A. Rous and Sarik Ghazarian and Sayan Ghosh and Sean Casey and Sebastian Bischoff and Sebastian Gehrmann and Sebastian Schuster and Sepideh Sadeghi and Shadi Hamdan and Sharon Zhou and Shashank Srivastava and Sherry Shi and Shikhar Singh and Shima Asaadi and Shixiang Shane Gu and Shubh Pachchigar and Shubham Toshniwal and Shyam Upadhyay and Shyamolima Shammie Debnath and Siamak Shakeri and Simon Thormeyer and Simone Melzi and Siva Reddy and Sneha Priscilla Makini and Soo-Hwan Lee and Spencer Torene and Sriharsha Hatwar and Stanislas Dehaene and Stefan Divic and Stefano Ermon and Stella Biderman and Stephanie Lin and Stephen Prasad and Steven Piantadosi and Stuart Shieber and Summer Misherghi and Svetlana Kiritchenko and Swaroop Mishra and Tal Linzen and Tal Schuster and Tao Li and Tao Yu and Tariq Ali and Tatsunori Hashimoto and Te-Lin Wu and Th{\'e}o Desbordes and Theodore Rothschild and Thomas Phan and Tianle Wang and Tiberius Nkinyili and Timo Schick and Timofei Kornev and Titus Tunduny and Tobias Gerstenberg and Trenton Chang and Trishala Neeraj and Tushar Khot and Tyler Shultz and Uri Shaham and Vedant Misra and Vera Demberg and Victoria Nyamai and Vikas Raunak and Vinay Venkatesh Ramasesh and vinay uday prabhu and Vishakh Padmakumar and Vivek Srikumar and William Fedus and William Saunders and William Zhang and Wout Vossen and Xiang Ren and Xiaoyu Tong and Xinran Zhao and Xinyi Wu and Xudong Shen and Yadollah Yaghoobzadeh and Yair Lakretz and Yangqiu Song and Yasaman Bahri and Yejin Choi and Yichi Yang and Sophie Hao and Yifu Chen and Yonatan Belinkov and Yu Hou and Yufang Hou and Yuntao Bai and Zachary Seid and Zhuoye Zhao and Zijian Wang and Zijie J. Wang and Zirui Wang and Ziyi Wu},
journal={Transactions on Machine Learning Research},
issn={2835-8856},
year={2023},
url={https://openreview.net/forum?id=uyTL5Bvosj},
note={Featured Certification}
}

@inproceedings{
agentbench,
title={{AgentBench}: Evaluating {LLM}s as Agents},
author={Xiao Liu and Hao Yu and Hanchen Zhang and Yifan Xu and Xuanyu Lei and Hanyu Lai and Yu Gu and Hangliang Ding and Kaiwen Men and Kejuan Yang and Shudan Zhang and Xiang Deng and Aohan Zeng and Zhengxiao Du and Chenhui Zhang and Sheng Shen and Tianjun Zhang and Yu Su and Huan Sun and Minlie Huang and Yuxiao Dong and Jie Tang},
booktitle={The Twelfth International Conference on Learning Representations},
year={2024},
url={https://openreview.net/forum?id=zAdUB0aCTQ}
}

@inproceedings{ctibench,
 author = {Alam, Md Tanvirul and Bhusal, Dipkamal and Nguyen, Le and Rastogi, Nidhi},
 booktitle = {Advances in Neural Information Processing Systems},
 doi = {10.52202/079017-1607},
 editor = {A. Globerson and L. Mackey and D. Belgrave and A. Fan and U. Paquet and J. Tomczak and C. Zhang},
 pages = {50805--50825},
 publisher = {Curran Associates, Inc.},
 title = {{CTIBench}: A Benchmark for Evaluating {LLM}s in Cyber Threat Intelligence},
 url = {https://proceedings.neurips.cc/paper_files/paper/2024/file/5acd3c628aa1819fbf07c39ef73e7285-Paper-Datasets_and_Benchmarks_Track.pdf},
 volume = {37},
 year = {2024}
}

@article{foundationsec_instruct,
  title={{Llama-3.1-FoundationAI-SecurityLLM-8B-Instruct} Technical Report},
  author={Weerawardhena, Sajana and Kassianik, Paul and Nelson, Blaine and Saglam, Baturay and Vellore, Anu and Priyanshu, Aman and Vijay, Supriti and Aufiero, Massimo and Goldblatt, Arthur and Burch, Fraser and Li, Ed and He, Jianliang and Kedia, Dhruv and Oshiba, Kojin and Yang, Zhuoran and Singer, Yaron and Karbasi, Amin},
  journal={arXiv preprint arXiv:2508.01059},
  year={2025},
  doi={10.48550/arXiv.2508.01059},
  url={https://arxiv.org/abs/2508.01059}
}

@article{foundationsec_reasoning,
  title={{Llama-3.1-FoundationAI-SecurityLLM-Reasoning-8B} Technical Report},
  author={Yang, Zhuoran and Li, Ed and He, Jianliang and Priyanshu, Aman and Saglam, Baturay and Kassianik, Paul and Weerawardhena, Sajana and Vellore, Anu and Nelson, Blaine and Javidnia, Neusha and Goldblatt, Arthur and Burch, Fraser and Zohary, Avi and Eisenman, Assaf and Sabbaghi, Mahdi and Vijay, Supriti and Dharssi, Rahim and Kedia, Dhruv and Oshiba, Kojin and Singer, Yaron and Karbasi, Amin},
  journal={arXiv preprint arXiv:2601.21051},
  year={2026},
  doi={10.48550/arXiv.2601.21051},
  url={https://arxiv.org/abs/2601.21051}
}

@misc{openai_gpt41,
  title={Introducing {GPT-4.1} in the {API}},
  author={{OpenAI}},
  year={2025},
  note={\url{https://openai.com/index/gpt-4-1/}}
}

@misc{openai_gpt5,
  title={Introducing {GPT-5} for Developers},
  author={{OpenAI}},
  year={2025},
  note={\url{https://openai.com/index/introducing-gpt-5-for-developers/}}
}

@article{gemma3,
  title={{Gemma 3} Technical Report},
  author={{Gemma Team}},
  journal={arXiv preprint arXiv:2503.19786},
  year={2025},
  url={https://arxiv.org/abs/2503.19786}
}

@misc{claudecode,
  title={{Claude Code}: An Agentic Coding Tool},
  author={Anthropic},
  year={2025},
  note={\url{https://docs.anthropic.com/en/docs/claude-code}}
}

@misc{karpathy_autoresearch,
  title={{autoresearch}: {AI} Agents Running Research on Single-{GPU} {nanochat} Training Automatically},
  author={Karpathy, Andrej},
  year={2026},
  note={\url{https://github.com/karpathy/autoresearch}}
}

@article{gcg,
  title={Universal and Transferable Adversarial Attacks on Aligned Language Models},
  author={Zou, Andy and Wang, Zifan and Carlini, Nicholas and Nasr, Milad and Kolter, J. Zico and Fredrikson, Matt},
  journal={arXiv preprint arXiv:2307.15043},
  year={2023}
}

@inproceedings{
autodan,
title={{AutoDAN}: Generating Stealthy Jailbreak Prompts on Aligned Large Language Models},
author={Xiaogeng Liu and Nan Xu and Muhao Chen and Chaowei Xiao},
booktitle={The Twelfth International Conference on Learning Representations},
year={2024},
url={https://openreview.net/forum?id=7Jwpw4qKkb}
}

@article{pair,
  title={Jailbreaking Black Box Large Language Models in Twenty Queries},
  author={Chao, Patrick and Robey, Alexander and Dobriban, Edgar and Hassani, Hamed and Pappas, George J. and Wong, Eric},
  journal={arXiv preprint arXiv:2310.08419},
  year={2023}
}

@article{evox,
  title={{EvoX}: Meta-Evolution for Automated Discovery},
  author={Liu, Shu and Agarwal, Shubham and others},
  journal={arXiv preprint arXiv:2602.23413},
  year={2026}
}

@article{adaevolve,
  title={{AdaEvolve}: Adaptive {LLM}-Driven Zeroth-Order Optimization},
  author={Cemri, Mert and Agrawal, Shubham and Gupta, Akshat and Liu, Shu and Cheng, Audrey and Mang, Qiuyang and Naren, Ashwin and Erdogan, Lutfi Eren and Sen, Koushik and Zaharia, Matei and Dimakis, Alex and Stoica, Ion},
  journal={arXiv preprint arXiv:2602.20133},
  year={2026}
}

@article{claudini,
  title={{Claudini}: Autoresearch Discovers State-of-the-Art Adversarial Attack Algorithms for {LLM}s},
  author={Panfilov, Alexander and Romov, Peter and Shilov, Igor and de Montjoye, Yves-Alexandre and Geiping, Jonas and Andriushchenko, Maksym},
  journal={arXiv preprint arXiv:2603.24511},
  year={2026}
}

@inproceedings{
opro,
title={Large Language Models as Optimizers},
author={Chengrun Yang and Xuezhi Wang and Yifeng Lu and Hanxiao Liu and Quoc V Le and Denny Zhou and Xinyun Chen},
booktitle={The Twelfth International Conference on Learning Representations},
year={2024},
url={https://openreview.net/forum?id=Bb4VGOWELI}
}

@inproceedings{
ape,
title={Large Language Models are Human-Level Prompt Engineers},
author={Yongchao Zhou and Andrei Ioan Muresanu and Ziwen Han and Keiran Paster and Silviu Pitis and Harris Chan and Jimmy Ba},
booktitle={The Eleventh International Conference on Learning Representations },
year={2023},
url={https://openreview.net/forum?id=92gvk82DE-}
}

@article{textgrad,
  title={{TextGrad}: Automatic ``Differentiation'' via Text},
  author={Yuksekgonul, Mert and Bianchi, Federico and Boen, Joseph and Liu, Sheng and Huang, Zhi and Guestrin, Carlos and Zou, James},
  journal={arXiv preprint arXiv:2406.07496},
  year={2024}
}

@inproceedings{mipro,
    title = "Optimizing Instructions and Demonstrations for Multi-Stage Language Model Programs",
    author = "Opsahl-Ong, Krista  and
      Ryan, Michael J  and
      Purtell, Josh  and
      Broman, David  and
      Potts, Christopher  and
      Zaharia, Matei  and
      Khattab, Omar",
    editor = "Al-Onaizan, Yaser  and
      Bansal, Mohit  and
      Chen, Yun-Nung",
    booktitle = "Proceedings of the 2024 Conference on Empirical Methods in Natural Language Processing",
    month = nov,
    year = "2024",
    address = "Miami, Florida, USA",
    publisher = "Association for Computational Linguistics",
    url = "https://aclanthology.org/2024.emnlp-main.525/",
    doi = "10.18653/v1/2024.emnlp-main.525",
    pages = "9340--9366"
}

@inproceedings{triggers,
    title = "Universal Adversarial Triggers for Attacking and Analyzing {NLP}",
    author = "Wallace, Eric  and
      Feng, Shi  and
      Kandpal, Nikhil  and
      Gardner, Matt  and
      Singh, Sameer",
    editor = "Inui, Kentaro  and
      Jiang, Jing  and
      Ng, Vincent  and
      Wan, Xiaojun",
    booktitle = "Proceedings of the 2019 Conference on Empirical Methods in Natural Language Processing and the 9th International Joint Conference on Natural Language Processing (EMNLP-IJCNLP)",
    month = nov,
    year = "2019",
    address = "Hong Kong, China",
    publisher = "Association for Computational Linguistics",
    url = "https://aclanthology.org/D19-1221/",
    doi = "10.18653/v1/D19-1221",
    pages = "2153--2162"
}

@inproceedings{autoprompt,
    title = "{A}uto{P}rompt: {E}liciting {K}nowledge from {L}anguage {M}odels with {A}utomatically {G}enerated {P}rompts",
    author = "Shin, Taylor  and
      Razeghi, Yasaman  and
      Logan IV, Robert L.  and
      Wallace, Eric  and
      Singh, Sameer",
    editor = "Webber, Bonnie  and
      Cohn, Trevor  and
      He, Yulan  and
      Liu, Yang",
    booktitle = "Proceedings of the 2020 Conference on Empirical Methods in Natural Language Processing (EMNLP)",
    month = nov,
    year = "2020",
    address = "Online",
    publisher = "Association for Computational Linguistics",
    url = "https://aclanthology.org/2020.emnlp-main.346/",
    doi = "10.18653/v1/2020.emnlp-main.346",
    pages = "4222--4235"
}

@article{tap,
  title={Tree of Attacks: Jailbreaking Black-Box {LLMs} Automatically},
  author={Mehrotra, Anay and Zampetakis, Manolis and Kassianik, Paul and Nelson, Blaine and Anderson, Hyrum and Singer, Yaron and Karbasi, Amin},
  journal={Advances in Neural Information Processing Systems},
  year={2024}
}

@inproceedings{bonjailbreak,
  title={Best-of-{N} Jailbreaking},
  author={Hughes, John and Price, Sara and Lynch, Aengus and Schaeffer, Rylan and Barez, Fazl and Koyejo, Sanmi and Sleight, Henry and Jones, Erik and Perez, Ethan and Sharma, Mrinank},
  booktitle={Advances in Neural Information Processing Systems},
  year={2025},
  url={https://proceedings.neurips.cc/paper_files/paper/2025/hash/69f3eb242c7c9df9ea2f2b66ea8b3c0f-Abstract-Conference.html}
}

@inproceedings{
capscaling,
title={Capability-Based Scaling Trends for {LLM}-Based Red-Teaming},
author={Alexander Panfilov and Paul Kassianik and Maksym Andriushchenko and Jonas Geiping},
booktitle={The Fourteenth International Conference on Learning Representations},
year={2026},
url={https://openreview.net/forum?id=1InFGGz1D5}
}

@inproceedings{
advreasoning,
title={Adversarial Reasoning at Jailbreaking Time},
author={Mahdi Sabbaghi and Paul Kassianik and George J. Pappas and Amin Karbasi and Hamed Hassani},
booktitle={Forty-second International Conference on Machine Learning},
year={2025},
url={https://openreview.net/forum?id=aWd7mL5U9Q}
}

@inproceedings{
evoprompt,
title={Connecting Large Language Models with Evolutionary Algorithms Yields Powerful Prompt Optimizers},
author={Qingyan Guo and Rui Wang and Junliang Guo and Bei Li and Kaitao Song and Xu Tan and Guoqing Liu and Jiang Bian and Yujiu Yang},
booktitle={The Twelfth International Conference on Learning Representations},
year={2024},
url={https://openreview.net/forum?id=ZG3RaNIsO8}
}

@inproceedings{promptbreeder,
author = {Fernando, Chrisantha and Banarse, Dylan and Michalewski, Henryk and Osindero, Simon and Rockt\"{a}schel, Tim},
title = {{PromptBreeder}: self-referential self-improvement via prompt evolution},
year = {2024},
publisher = {JMLR.org},
booktitle = {Proceedings of the 41st International Conference on Machine Learning},
articleno = {541},
numpages = {64},
location = {Vienna, Austria},
series = {ICML'24}
}

@article{
helm,
title={Holistic Evaluation of Language Models},
author={Percy Liang and Rishi Bommasani and Tony Lee and Dimitris Tsipras and Dilara Soylu and Michihiro Yasunaga and Yian Zhang and Deepak Narayanan and Yuhuai Wu and Ananya Kumar and Benjamin Newman and Binhang Yuan and Bobby Yan and Ce Zhang and Christian Alexander Cosgrove and Christopher D Manning and Christopher Re and Diana Acosta-Navas and Drew Arad Hudson and Eric Zelikman and Esin Durmus and Faisal Ladhak and Frieda Rong and Hongyu Ren and Huaxiu Yao and Jue WANG and Keshav Santhanam and Laurel Orr and Lucia Zheng and Mert Yuksekgonul and Mirac Suzgun and Nathan Kim and Neel Guha and Niladri S. Chatterji and Omar Khattab and Peter Henderson and Qian Huang and Ryan Andrew Chi and Sang Michael Xie and Shibani Santurkar and Surya Ganguli and Tatsunori Hashimoto and Thomas Icard and Tianyi Zhang and Vishrav Chaudhary and William Wang and Xuechen Li and Yifan Mai and Yuhui Zhang and Yuta Koreeda},
journal={Transactions on Machine Learning Research},
issn={2835-8856},
year={2023},
url={https://openreview.net/forum?id=iO4LZibEqW},
note={Featured Certification, Expert Certification}
}

@inproceedings{hover,
    title = "{H}o{V}er: A Dataset for Many-Hop Fact Extraction And Claim Verification",
    author = "Jiang, Yichen  and
      Bordia, Shikha  and
      Zhong, Zheng  and
      Dognin, Charles  and
      Singh, Maneesh  and
      Bansal, Mohit",
    editor = "Cohn, Trevor  and
      He, Yulan  and
      Liu, Yang",
    booktitle = "Findings of the Association for Computational Linguistics: EMNLP 2020",
    month = nov,
    year = "2020",
    address = "Online",
    publisher = "Association for Computational Linguistics",
    url = "https://aclanthology.org/2020.findings-emnlp.309/",
    doi = "10.18653/v1/2020.findings-emnlp.309",
    pages = "3441--3460"
}

@inproceedings{papillon,
    title = "{PAPILLON}: Privacy Preservation from {I}nternet-based and Local Language Model Ensembles",
    author = "Siyan, Li  and
      Raghuram, Vethavikashini Chithrra  and
      Khattab, Omar  and
      Hirschberg, Julia  and
      Yu, Zhou",
    editor = "Chiruzzo, Luis  and
      Ritter, Alan  and
      Wang, Lu",
    booktitle = "Proceedings of the 2025 Conference of the Nations of the Americas Chapter of the Association for Computational Linguistics: Human Language Technologies (Volume 1: Long Papers)",
    month = apr,
    year = "2025",
    address = "Albuquerque, New Mexico",
    publisher = "Association for Computational Linguistics",
    url = "https://aclanthology.org/2025.naacl-long.173/",
    doi = "10.18653/v1/2025.naacl-long.173",
    pages = "3371--3390",
    ISBN = "979-8-89176-189-6"
}

@inproceedings{ifbench,
    title={Generalizing Verifiable Instruction Following},
    author={Valentina Pyatkin and Saumya Malik and Victoria Graf and Hamish Ivison and Shengyi Huang and Pradeep Dasigi and Nathan Lambert and Hannaneh Hajishirzi},
    booktitle={The Thirty-ninth Annual Conference on Neural Information Processing Systems Datasets and Benchmarks Track},
    year={2026},
    url={https://openreview.net/forum?id=yfYgwjj5F8}
}

@inproceedings{livebench,
    title={{LiveBench}: A Challenging, Contamination-Limited {LLM} Benchmark},
    author={Colin White and Samuel Dooley and Manley Roberts and Arka Pal and Benjamin Feuer and Siddhartha Jain and Ravid Shwartz-Ziv and Neel Jain and Khalid Saifullah and Sreemanti Dey and Shubh-Agrawal and Sandeep Singh Sandha and Siddartha Venkat Naidu and Chinmay Hegde and Yann LeCun and Tom Goldstein and Willie Neiswanger and Micah Goldblum},
    booktitle={The Thirteenth International Conference on Learning Representations},
    year={2025},
    url={https://openreview.net/forum?id=sKYHBTAxVa}
}

@misc{aime,
      title={Beyond Benchmarks: {MathArena} as an Evaluation Platform for Mathematics with {LLM}s},
      author={Jasper Dekoninck and Nikola Jovanović and Tim Gehrunger and Kári Rögnvaldsson and Ivo Petrov and Chenhao Sun and Martin Vechev},
      year={2026},
      eprint={2605.00674},
      archivePrefix={arXiv},
      primaryClass={cs.CL},
      url={https://arxiv.org/abs/2605.00674},
}
}

\clearpage

\appendix

\section{System Implementation Details}
\label{app:system-details}

This appendix gives the technical details that are summarized at a higher level in Section~\ref{sec:system}.

\subsection{Runtime and Task Workspaces}
\label{app:runtime-workspaces}

\FAPO{} is implemented as a reusable evaluation runtime plus tenant-local pipeline definitions.
The reusable runtime lives under \code{src/hephaestus/}.
It contains typed config objects, dataset loading, prompt rendering, provider adapters, LangGraph chain loading, scoring, run-artifact writing, progress tracking, storage helpers, and post-hoc failure attribution.
The tenant layer lives under \code{tenants/<tenant\_id>/}.
It contains the task chain, prompt variants, scorer implementation, dataset conversion scripts, local configs, data contracts, and iteration history for a single task.

The runtime boundary is the eval config.
The config is parsed into an \code{EvalConfig} with fields for \code{tenant\_id}, \code{provider}, \code{provider\_settings}, \code{dataset\_path}, \code{scoring\_profile}, \code{output\_dir}, optional \code{max\_workers}, optional \code{run\_id}, and a \code{ChainConfig}.
The \code{ChainConfig} contains a tenant chain module path, a factory function name, and an arbitrary chain-local config dictionary.
In practice this dictionary carries \code{prompt\_paths} and task parameters such as retrieval depth.
Before an eval starts, the runner checks that the dataset, chain module, and every configured prompt file exist.

The end-to-end control flow is:

\begin{enumerate}
    \item \code{load\_eval\_config} reads the JSON config and validates the provider, dataset path, chain path, chain function, and concurrency settings.
    \item \code{load\_cases} reads unified JSONL cases into \code{EvalCase} records with \code{case\_id}, \code{task\_type}, \code{context}, \code{expected}, \code{metadata}, and optional prompt-template fields.
    \item \code{load\_tenant\_scorer} dynamically imports the tenant scorer class named in \code{scoring\_profile.scorer}.
    \item \code{build\_provider\_client} constructs a provider adapter for OpenAI, SageMaker, or Baseten-compatible inference.
    \item \code{load\_chain\_factory} imports the tenant chain factory and calls \code{build\_chain(provider, config)} to obtain a compiled LangGraph graph.
    \item The runner streams every case through the graph, scores the resulting state, updates progress, and writes durable run artifacts.
\end{enumerate}

This design makes the task model interchangeable behind a small \code{ProviderClient.generate(messages)} interface, while keeping task-specific logic outside the core package.
Single-node tenants such as AIME and CTIBench-RCM instantiate a one-node LangGraph chain.
The HotpotQA tenant instantiates a six-node chain with BM25 retrieval, summarization, follow-up-query generation, a second retrieval hop, a second summary, and final answer generation.
Both forms use the same runner and scorer contract.

\subsection{Chains and Pipeline-Aware Scoring}
\label{app:chains-scoring}

Evaluation targets are LangGraph \code{StateGraph} objects compiled into executable chains.
The chain factory signature is fixed:
\code{build\_chain(provider: ProviderClient, config: Dict[str, Any]) -> CompiledGraph}.
The factory may build a typed graph with \code{StateGraph(ChainState)} or an equivalent dictionary-state graph, but the graph must preserve the state fields in Table~\ref{tab:chainstate}.
The runner initializes these fields for every case and also adds an internal worker index when concurrent evaluation is enabled.

\begin{table}[!ht]
\centering
\small
\renewcommand{\arraystretch}{1.3}
\begin{tabular}{@{}lll@{}}
\toprule
\textbf{Field} & \textbf{Type} & \textbf{Description} \\
\midrule
\code{context} & \code{Dict[str, str]} & Input from the evaluation case \\
\code{output\_text} & \code{str} & Final chain output for scoring \\
\code{step\_outputs} & \code{Dict[str, str]} & Intermediate step results, keyed by node name \\
\code{diagnostics} & \code{List[str]} & Debug traces and warnings \\
\bottomrule
\end{tabular}
\caption{The \code{ChainState} protocol. Tenant chains may type this state explicitly or use an equivalent dictionary state, and may extend it with additional fields.}
\label{tab:chainstate}
\end{table}

\FAPO{} provides a node factory, \code{make\_llm\_node}, for ordinary LLM calls.
The factory reads a prompt template once at chain-construction time.
At runtime the node builds a render context by merging the case \code{context} with prior step outputs under keys of the form \code{steps.<name>.output}.
It then renders \code{\$\{...\}} placeholders, converts templates with \code{System:} and \code{User:} sections into chat messages, calls \code{provider.generate}, optionally applies an output parser, and returns a state update.
The update sets \code{output\_text} to the node output, writes the same value into \code{step\_outputs[output\_key]}, and appends any missing-placeholder diagnostics.

Custom nodes use the same state-update contract.
For example, the HotpotQA retrieval node reads either a case-context key or a previous step output, queries an in-process BM25 index, and writes formatted passages into \code{step\_outputs}.
The second-hop query-generation prompt can reference the first summary as \code{\$\{steps.summarize\_hop1.output\}}, and the answer-generation prompt can reference retrieval and summary outputs from both hops.
Because every node writes a named output, \FAPO{} can inspect the pipeline as a sequence of typed intermediate artifacts rather than as a single opaque final string.

Evaluation uses \code{chain.stream(initial\_state)} rather than a single blocking invoke.
After each streamed LangGraph chunk, the runner merges the node update into the final state and records \code{step\_timings} as \code{[node\_name, elapsed\_seconds]} pairs.
If a chain raises an exception, the runner records the exception in \code{diagnostics} and scores the case with an empty output so that the remaining cases can continue.
If a chain never sets \code{output\_text}, the runner emits a warning and scores an empty final answer.

Scoring is also pipeline-aware.
Tenant scorers subclass \code{src.hephaestus.scoring.scorer.Scorer}.
Every scorer implements \code{validate\_case} and \code{score\_case}; chain-aware scorers may override \code{score\_pipeline\_case(case, step\_outputs, scoring\_profile, output\_text)}.
The default implementation scores the final output, while HotpotQA explicitly scores the \code{answer} step when present.
The runtime validates that scorer payloads contain a finite \code{composite\_score} in \([0,100]\) and a numeric \code{score\_breakdown}.
Benchmark-specific scorers can expose extra metrics, such as exact match, F1, answer-format validity, point totals, or LLM-judge equivalence, while preserving one comparable optimization objective.

\subsection{Run Artifacts and Failure Attribution}
\label{app:artifacts-attribution}

Each eval writes a self-contained output directory.
\code{run\_config.json} records the resolved provider settings, dataset path, scoring profile, max-worker setting, run id, and chain config used for the run.
\code{results.jsonl} stores one \code{EvalCaseResult} per case, including \code{case\_id}, \code{task\_type}, \code{diagnostics}, \code{score\_breakdown}, \code{composite\_score}, \code{output\_text}, \code{step\_outputs}, and \code{step\_timings}.
\code{progress.json} is written atomically during execution and contains run status, completed and in-flight case ids, average composite score, score-breakdown averages, and point-weighted averages when the scorer reports earned and possible points.
\code{summary.md} reports aggregate scores, score-breakdown averages, per-step timing statistics, and, when failures include step-level evidence, an automatic step-attribution table.

The attribution implementation is deliberately lightweight and deterministic before Claude performs deeper analysis.
\code{attribute\_failures} loads \code{results.jsonl}, filters cases below a score threshold, and assigns failures to likely chain steps using named heuristics.
Retrieval and search steps are recognized by step names containing retrieval-like terms.
For those steps, the analyzer computes question-to-output token overlap and classifies retrieval quality as hit, partial, or miss.
It also detects intermediate steps with empty recorded results, cascading failures caused by an early empty step, final-answer format failures where the expected answer appears with extra text, and a low-confidence final-step fallback.
\code{summarize} then partitions failures into prompt-addressable and structurally-addressable buckets and reports counts by confidence and retrieval tier.

For long agentic traces, \FAPO{} can attach richer evidence without changing the scoring contract.
\code{trace\_loader} joins a case row from \code{results.jsonl} with an optional Inspect \code{.eval} log and produces a compact trajectory containing turns, tool calls, tool results, errors, token counts, wall-clock time, expected answer, and final output.
If no Inspect log is present, it degrades to a trajectory synthesized from \code{step\_outputs} and \code{step\_timings}.
The optimization layer uses these digests only as evidence; step attribution remains the component that emits failure clusters and recommended optimization levels.

\subsection{Tenant Isolation}
\label{app:tenant-isolation}

Each tenant is self-contained.
The expected directory layout includes \code{source\_artifacts/} for protected raw inputs, \code{datasets/} for local derived JSONL caches, \code{code/} for conversion and scorer helpers, \code{tests/} for tenant-specific assumptions, \code{chains/} for baseline and structural variants, \code{prompts/} for prompt variants, \code{configs/} for ephemeral eval configs, \code{storage/config.json} for customer-data synchronization, \code{docs/} for operating contracts, \code{evals/} for local run outputs, and \code{reports/} for local analysis notes.
Core eval code consumes only unified JSONL cases at runtime; tenant-specific raw-data adapters are offline conversion scripts under \code{tenants/<tenant\_id>/code/}.

The evaluation config is the first enforcement point: dataset paths, chain paths, prompt paths, scorer modules, and output directories are tenant-local.
Customer raw and derived artifacts are pulled, pushed, or removed through \code{python -m hephaestus.cli customer-data}, with canonical storage described by the tenant's tracked \code{storage/config.json}.
Local eval configs and run outputs are treated as ephemeral workspace artifacts, while tenant docs such as \code{data-contract.md}, \code{prompt-contract.md}, \code{eval-operations.md}, \code{iteration-playbook.md}, \code{change-log.md}, and \code{iteration-memory.jsonl} are checked in as the operational contract for optimization.

During optimization, Claude reads the tenant's iteration playbook before editing and emits a scope contract that lists allowed optimization levels and forbidden changes.
Prompt variants are immutable: each iteration creates a new numbered file.
Structural variants are also cloned rather than edited in place, live under \code{chains/variants/}, include metadata describing the parent chain and hypothesis, and must obtain prompt paths from config rather than hardcoding tenant paths.
Every candidate prompt, parameter, or chain variant is checked against the scope contract before evaluation.
The variant-reviewer independently repeats the check and blocks cross-tenant paths, cross-tenant imports, copied examples or labels from another tenant, placeholder drift, data leakage, scorer incompatibility, state-protocol violations, and unsafe imports.
Thus isolation is enforced by directory layout, config-local paths, immutable variant conventions, optimizer self-checks, and reviewer validation; it is a workspace boundary rather than an operating-system sandbox.

\clearpage

\section{Optimized Prompt Variants}
\label{app:prompts}

Optimized prompts differ by model, even for the same task.
We show CTIBench-RCM prompts for each model and the HotpotQA answer-generation prompt before and after optimization.

\subsection{CTIBench-RCM: Baseline (variant-001, all models)}

\begin{lstlisting}[basicstyle=\ttfamily\scriptsize, breaklines=true, frame=single, xleftmargin=4pt, xrightmargin=4pt]
System: You are a cybersecurity expert specializing in
vulnerability analysis and weakness classification.

Analyze the following CVE description and map it to the
appropriate CWE. Provide a brief justification for your
choice. Ensure the last line of your response contains
only the CWE ID.

User: ${description}
\end{lstlisting}

\subsection{CTIBench-RCM: \GPTFive{} Best (variant-029, 76.1\% test)}

The best \GPTFive{} prompt adds NVD convention rules for specific CWE confusion pairs.
The prompt grows from 4 lines to 23:

\begin{lstlisting}[basicstyle=\ttfamily\scriptsize, breaklines=true, frame=single, xleftmargin=4pt, xrightmargin=4pt]
System: You are a cybersecurity expert specializing in
vulnerability analysis and weakness classification.

Analyze the following CVE description and map it to the
appropriate CWE. Provide a brief justification for your
choice. Ensure the last line of your response contains
only the CWE ID.

When selecting a CWE, follow NVD mapping conventions:
- Buffer overflows (stack/heap/unspecified) -> CWE-787,
  not CWE-121 or CWE-122.
- Command injection -> CWE-77, not CWE-78, unless the
  description explicitly describes OS-level commands.
- Hardcoded credentials -> CWE-798.
- DoS through malformed input -> CWE-404 when there is
  no indication of memory corruption.
- Weak crypto -> CWE-327.  Missing authz -> CWE-862.
- Integer overflow -> CWE-190.  NULL deref -> CWE-476.
- Use-after-free -> CWE-416.
- Observable timing/side-channel -> CWE-203.
- Info exposure through error messages -> CWE-209.

Common mistakes to avoid:
- Do NOT use CWE-20 as a catch-all.
- Focus on root cause, not impact or attack vector.

User: ${description}
\end{lstlisting}

\subsection{CTIBench-RCM: \FoundationSecEightBInstruct{} Best (variant-037, 71.0\% test)}

For the Instruct model, added rules hurt format extraction.
The best prompt is two lines---a 2$\times$ reduction from baseline:

\begin{lstlisting}[basicstyle=\ttfamily\scriptsize, breaklines=true, frame=single, xleftmargin=4pt, xrightmargin=4pt]
System: You are a CWE classification expert following NVD
mapping conventions. Given a CVE description, identify the
root cause CWE. Output the CWE ID on the last line.

User: ${description}
\end{lstlisting}

\subsection{CTIBench-RCM: \FoundationSecEightBReasoning{} Best (variant-072, 73.0\% test)}

The Reasoning model's best prompt is almost the same as the Instruct prompt.
The phrase ``standard NVD abstraction level'' accounts for +2.9\,pp in ablation:

\begin{lstlisting}[basicstyle=\ttfamily\scriptsize, breaklines=true, frame=single, xleftmargin=4pt, xrightmargin=4pt]
System: You are a CWE classification expert. Map the CVE
description to the most appropriate CWE following NVD
mapping conventions. Use the standard NVD abstraction
level. Output the CWE ID on the last line.

User: ${description}
\end{lstlisting}

\subsection{HotpotQA: Answer Generation Before and After}

The HotpotQA baseline uses a bare DSPy-format prompt.
Variant-003 adds brevity rules, a must-answer rule, and format guidance.
These changes target near-miss and abstention failures.

\paragraph{Baseline (variant-001, 39.22\% val EM):}

\begin{lstlisting}[basicstyle=\ttfamily\scriptsize, breaklines=true, frame=single, xleftmargin=4pt, xrightmargin=4pt]
System: Your input fields are:
1. `question` (str):
2. `summary_1` (str):
3. `summary_2` (str):
Your output fields are:
1. `reasoning` (str):
2. `answer` (str):
[...]
In adhering to this structure, your objective is:
  Given the fields `question`, `summary_1`, `summary_2`,
  produce the fields `answer`.
\end{lstlisting}

\paragraph{Optimized (variant-003, 70.3\% val EM):}

\begin{lstlisting}[basicstyle=\ttfamily\scriptsize, breaklines=true, frame=single, xleftmargin=4pt, xrightmargin=4pt]
System: You answer multi-hop questions with the SHORTEST
possible answer.

CRITICAL RULES:
1. MUST ALWAYS provide an answer. NEVER say "unknown",
   "none", "N/A", or "not enough information".
2. If summaries contain partial info, use what you have
   to make your best inference.
3. If the question asks for a comparison and you only
   have data for one entity, answer with that entity.

ANSWER FORMAT RULES (follow EXACTLY):
- Output ONLY the entity name, number, date, or yes/no.
- NEVER output a full sentence as the answer.
- For yes/no questions: "yes" or "no" (lowercase).
- For "who": just the full name (e.g., "James Cameron").
- For "when": just the date (e.g., "1066").
- Copy names EXACTLY as spelled in the summaries.
- Use SINGULAR form when the question asks "what".
\end{lstlisting}

\section{CTIBench-RCM Full Variant Progression}
\label{app:progression}

Table~\ref{tab:gpt5-progression} shows the \GPTFive{} variant progression on the dev set.
The agent tested 31 variants, with scores ranging from 76.3\% to 85.6\%.
Early variants tried broad abstraction rules and regressed.
Variant-005 introduced NVD rules for specific CWE confusion pairs and jumped +4.1\,pp.
Subsequent variants refined the rule set, with diminishing returns past variant-026.

\begin{table}[h]
\centering
\scriptsize
\renewcommand{\arraystretch}{1.3}
\begin{tabular}{@{}clc@{}}
\toprule
\textbf{Var.} & \textbf{Strategy} & \textbf{Dev EM} \\
\midrule
001 & Baseline (simple classify prompt) & 78.6 \\
002 & Prefer parent CWE + examples & 76.3 \\
005 & NVD rules (787, 77, 404, 798) & 82.7 \\
012 & CWE-787 + CWE-798 only & 80.9 \\
020 & +\,CWE-327/862 rules & 83.2 \\
022 & +\,CWE-190/476/416 rules & 83.8 \\
026 & +\,negative examples + CWE-203/209 & 84.4 \\
029 & Refined CWE-404 wording & \textbf{85.6} \\
031 & +\,CWE-287/401 rules (regressed) & 83.8 \\
\bottomrule
\end{tabular}
\caption{\GPTFive{} variant progression on CTIBench-RCM dev set (173 cases). Selected variants shown; full history in the tenant change log.}
\label{tab:gpt5-progression}
\end{table}

\end{document}